\documentclass[aps,prl,twocolumn,superscriptaddress,nofootinbib]{revtex4-2}

\usepackage{amsmath}
\usepackage{amssymb}
\usepackage{algorithm}
\usepackage{algpseudocode}
\usepackage{bm}
\usepackage{graphicx}
\usepackage[colorlinks=true,allcolors=blue]{hyperref}
\usepackage{comment}

\begin{document}

% Exclude the main article and bibliography from the supplemental contents.
\addtocontents{toc}{\protect\setcounter{tocdepth}{-1}}

\title{Universal Scaling of Conditional Mutual Information in Classical Stochastic Dynamics}

\author{Songlei Wang}
\affiliation{School of Physics, Peking University, Beijing 100871, China}
\author{Chenguang Liang}
\affiliation{School of Physics, Peking University, Beijing 100871, China}
\author{Zhi-Cheng Yang}
\email{zcyang19@pku.edu.cn}
\affiliation{School of Physics, Peking University, Beijing 100871, China}
\affiliation{Center for High Energy Physics, Peking University, Beijing 100871, China}

\date{\today}

\begin{abstract}
Conditional mutual information (CMI) characterizes dependence between two regions conditioned on an intervening buffer. Its characteristic decay length, the Markov length, has recently been identified as a diverging length scale at mixed-state phase transitions. Here, we investigate CMI dynamics in one-dimensional classical stochastic models and propose a universal scaling form for systems whose Markov length diverges with time. We derive this form in the inactive phase of parity-conserving directed percolation, where domain-wall coarsening generates a growing length scale and a self-similar probability distribution. We introduce a Hankel-based truncation scheme that enables matrix-product-state simulations at long times and over large spatial regions. Using this method, we verify the proposed scaling both in the inactive phase and at the nonequilibrium critical point, where the dynamical critical exponent governs Markov-length growth. More broadly, the same scaling form captures cases in which collective inference sets the diverging Markov length. Despite these distinct mechanisms, the CMI between fixed microscopic regions exhibits a common scaling dimension of two. Finally, we extend this framework to temporal CMI, establishing an analogous relation between the strength of conditional dependence in time domain and its characteristic memory time.
\end{abstract}

\maketitle

\begin{figure}[t]
    \centering
    \includegraphics[width=0.85\columnwidth]{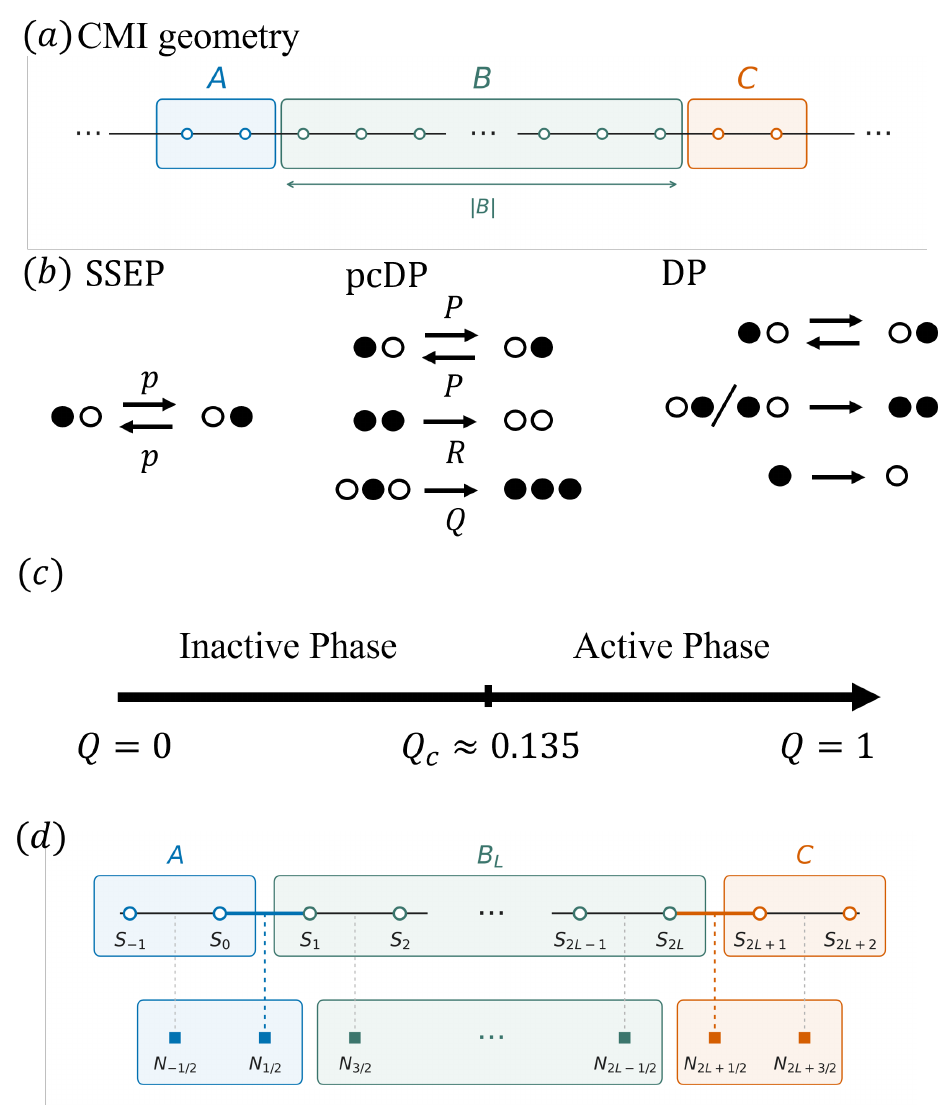}
    \caption{
    (a) Tripartition geometry for computing the CMI. Two fixed-size regions $A$ and $C$ are separated by a buffer region $B$ of variable length, all embedded in an infinite one-dimensional chain.
    (b) Allowed stochastic updates in the three models studied in this work: from left to right, the symmetric simple exclusion process (SSEP), parity-conserving directed percolation (PCDP), and directed percolation (DP). Filled and open circles represent occupied and empty sites, respectively, while arrows indicate the allowed transitions.
    (c) Phase diagram of the PCDP model as a function of branching rate $Q$, at fixed $P=0.50$ and $R=0.30$. The transition point $Q_c \approx 0.135$ was estimated using Monte Carlo simulations~\cite{SupplementalMaterial}.
    (d) Mapping between spin and domain-wall variables. Circles represent spin variables, while squares
    represent bond variables.
    }
    \label{fig:setup}
\end{figure}

\emph{Introduction---} Recent advances in quantum technologies, particularly the ability to control and engineer dissipation, have broadened the study of phases of matter and phase transitions beyond thermal Gibbs states to encompass generic mixed states~\cite{topoquantmemory,TOfiniteT,coser2019classification,openRMP,sang2024mixed,sang2025localreversibility,cmigapphase}. A key distinction is that generic mixed state phase transitions are often invisible to conventional correlation functions, where a diverging correlation length is responsible for singularities in such observables. Instead, intrinsic information-theoretic quantities have emerged as central tools for characterizing mixed-state phases and their transitions~\cite{QIgibbs,FB1,FB2,SWSSBreview}.
One particularly useful quantity is the conditional mutual information (CMI)~\cite{hayden2004structure,brown2012quantum,fawzi2015conditional} between two spatially separated regions $A$ and $C$, conditioned on an intervening buffer region $B$ [see Fig.~\ref{fig:setup}(a)]:
$I(A:C|B)=I(A:BC)-I(A:B)=H(AB)+H(BC)-H(B)-H(ABC)$, where $H(X)$ denotes the Shannon entropy of a classical probability distribution or the von Neumann entropy of a quantum state. The CMI quantifies the residual correlations between $A$ and $C$ after conditioning on $B$. Its exponential decay with the buffer size characterizes an approximate Markov property, with the Markov length $\xi_M$ setting the characteristic decay scale~\cite{kato2019thermal,sang2025stability,chen2025locally,CMIproofprx}. Importantly, a finite Markov length along an evolution path guarantees the existence of a short-depth quasi-local reverse quantum channel, which defines an equivalence relation between two mixed states in the same phase~\cite{sang2025localreversibility,sang2025stability}. Remarkably, the Markov length has been shown to diverge at critical points separating distinct mixed-state phases, including the error-correction threshold of the toric code~\cite{sang2025stability,noisetoriccodeCMI} and nonequilibrium critical points in certain classical stochastic models~\cite{lloyd2025,groverdp,hauser2026strong,iadecola2026local}.

In fact, a Markov length that diverges with time is rather ubiquitous in a {\it dynamical} setting \cite{lee2024universal,zhang2025information,2610CMIRG,Levin2610markovlength,MLdissipation}. For instance, it has been shown that $\xi_M(t) \propto t$ in a one-dimensional symmetric simple exclusion process (SSEP)~\cite{hauser2026strong}, and $\xi_M(t) \propto e^t$ following a thermal quench from an ordered state to an infinite-temperature state~\cite{lloyd2025}. In both cases, the growing Markov length reflects an emergent length scale associated with global inference, which increases as the system evolves. However, the scaling behavior of CMI in systems with a {\it dynamically} diverging Markov length has so far been largely characterized through heuristic scaling collapses.
Here, we establish a universal scaling form for CMI in systems whose Markov length diverges with time, valid in the regime of fixed finite $|A|, |C| \ll |B|$:
\begin{equation}
I_t(A:C|B)=\frac{1}{\ell(t)^2}
f\left(\frac{|B|}{\ell(t)}\right),
\label{eq:intro}
\end{equation}
where $\ell(t)$ is a growing characteristic length scale and $f$ is a model-dependent scaling function. When $f(x)\sim e^{-x}$ for large $x$, $\ell(t)$ can be identified with the Markov length $\xi_M(t)$. We derive this scaling form explicitly in a one-dimensional classical stochastic model belonging to the parity-conserving directed percolation (PCDP) universality class. In its inactive (absorbing) phase, particles undergo pair annihilation via domain-wall coarsening, which leads to a domain-wall distribution that is self-similar in time. We show that the Markov length is set by the average domain-wall separation and grows as $\xi_M(t)\sim t^{1/2}$, in agreement with our numerical results. At the critical point, we find that CMI obeys the same scaling form, with $\xi_M(t)\sim t^{1/z}$ governed by the growing dynamical correlation length. We further extend the analysis to CMI in the time domain, where the Markov length is interpreted as the memory time. By combining tensor-network simulations with Gaussian fluctuating hydrodynamics calculations, we demonstrate that the temporal CMI in a one-dimensional SSEP also follows the above scaling form. 

Numerically resolving this scaling form across different stochastic processes requires accurate finite-region probabilities at long times and for large system sizes. To this end, we develop a novel Hankel-based truncation scheme tailored to matrix-product-state (MPS) representations of classical probability distributions, which uses finite-word probabilities, rather than Schmidt weights, to guide bond-dimension truncation. Benchmarking the method in a directed percolation (DP) model, we find that it extends the time window over which small CMI tails can be reliably resolved from $t\simeq 900$ with conventional canonical compression to $t\simeq 1800$. In the PCDP model, this improved numerical accuracy is essential for reliably extracting the CMI scaling form.

\emph{Setup---}
We study the dynamics of CMI for a system of hard-core particles under classical stochastic processes in an infinite one-dimensional chain. Fig.~\ref{fig:setup}(b) summarizes the dynamical moves of all stochastic models studied in this work. 
The classical probability distribution at time $t$ can be written as a state vector $|P_t\rangle
=\sum_{\boldsymbol{s}} P_t(\boldsymbol{s})|\boldsymbol{s}\rangle$, with $P_t({\bm s})$ encoding the marginal probability of particle configuration ${\bm s}$ in a given spatial region.
Starting from an initial state $|P_0\rangle$, the state evolves according to the stochastic channel in Fig.~\ref{fig:setup}(b) $|P_t\rangle = \mathcal{E}_t |P_0\rangle$. Due to translational invariance, 
we can represent the state vector $|P_t\rangle$ as an 
infinite matrix-product state (iMPS)~\cite{temme2010stochastic,scramblinghydro,lloyd2025}. In particular, the
probability of observing a configuration
$(s_1,\ldots,s_n)$ on a consecutive $n$-site subsystem of an infinite chain at time $t$ is
\begin{equation}
    P_t(s_1,\ldots,s_n)
    =
    \langle\mathcal L_t|
    A_t^{s_1}\cdots A_t^{s_n}
    |\mathcal R_t\rangle .
    \label{eq:pmps}
\end{equation}
Here $A_t^s$ is the matrix associated with the local state $s$,
while $\langle\mathcal L_t|$ and $|\mathcal R_t\rangle$ encode the
two semi-infinite environments. They are the left and right fixed
points of the transfer matrix $T_t=\sum_s A_t^s$, normalized such that
$\langle\mathcal L_t|\mathcal R_t\rangle=1$ \cite{SupplementalMaterial}.
From the MPS representation, we compute the CMI using the perfect sampling method developed in Refs.~\cite{ferris2012perfect,lloyd2025}. Hereafter, we fix the size of $A$ and $C$ to contain either one or two sites, while varying the size of the buffer $B$.

\emph{Hankel truncation scheme---}
As in conventional MPS simulations, the bond dimension typically grows during time evolution, necessitating repeated truncation to keep the computation tractable while retaining the information encoded in the virtual bond space that is essential for accurately computing the observables of interest. For quantum states, retaining the largest Schmidt values yields the optimal fixed-rank approximation in the state-vector $2$-norm~\cite{schollwoeck2011}. For an iMPS representing a classical probability distribution, however, the natural observables of interest are finite-configuration probabilities, as defined in Eq.~\eqref{eq:pmps}, which are obtained through single-layer MPS contractions (see End Matter). The importance of a given virtual-space direction is therefore determined not only by its canonical weight, but also by how it can be accessed from $|\mathcal R_t\rangle$ and probed by $\langle\mathcal L_t|$ through finite sequences of local configurations. This distinction motivates an alternative truncation criterion that ranks virtual directions by their contribution to
finite-configuration probabilities directly, which we introduce in this work.

To formulate this criterion, let $w=(s_1,\ldots,s_{|w|})$ denote a finite sequence of local particle configurations of length $|w|$, and introduce the shorthand notation $A_t^w\equiv A_t^{s_1}\cdots A_t^{s_{|w|}}$. Partitioning $w$ into two consecutive sequences $u$ and $v$, such that $w=uv$, we define the Hankel matrix~\cite{carlyle1971,fliess1974,balle2014forward,balle2019}
\begin{equation}
H_t(u,v)
=
\underbrace{\langle\mathcal L_t|A_t^u}_{\text{prefix readout}}
\underbrace{A_t^v|\mathcal R_t\rangle}_{\text{suffix state}}
=
P_t(uv),
\end{equation}
whose rows and columns are indexed by all possible finite sequences $u$ and $v$, respectively. Each matrix element gives the probability of the concatenated configuration $uv$, with the contributions from the two semi-infinite environments incorporated through $\langle\mathcal L_t|$ and $|\mathcal R_t\rangle$. 
The Hankel truncation scheme compresses the singular value spectrum of this Hankel matrix
$H_t$, retaining the virtual subspace associated with its
largest singular values and hence the dominant contributions to finite-sequence marginal probabilities~\cite{moore1981,hankel1980,glover1984,antoulas2005,
hankelmpo,balle2019}. In practice, explicitly constructing the infinite-dimensional Hankel matrix is unnecessary; the implementation of the truncation scheme is detailed in the Supplemental Material (SM) \cite{SupplementalMaterial}.

\begin{figure}[t]
    \centering
    \includegraphics[width=0.85\columnwidth]{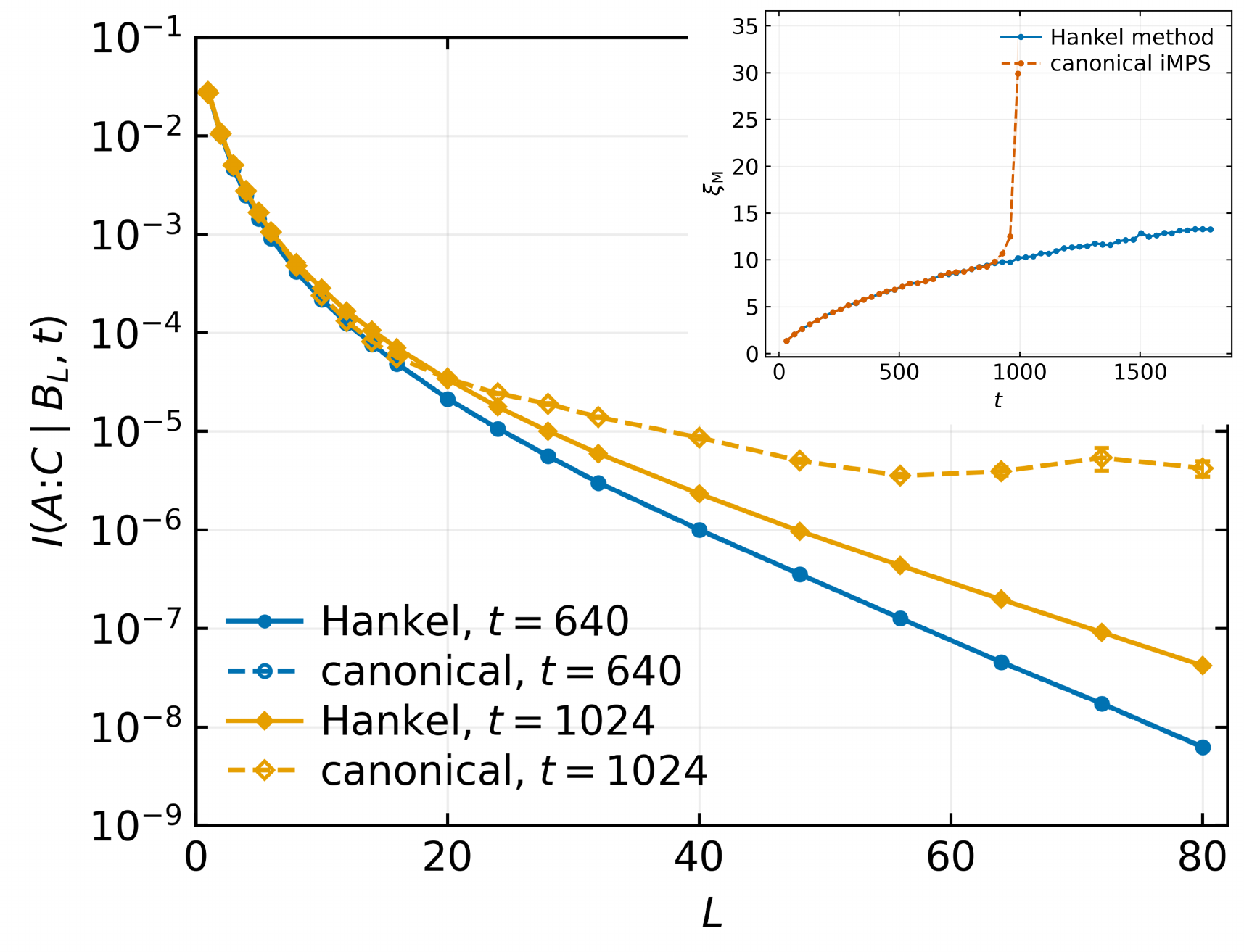}
    \caption{Comparison of Hankel-based truncation and canonical
    iMPS compression in critical DP.
    The main panel shows CMI profiles at $t=640$ and $1024$, computed using the same maximal bond dimension $\chi_{\rm max}=180$.
    The two methods agree at $t=640$, whereas at $t=1024$ the
    canonical truncation scheme leads to an artificial plateau at large buffer sizes, while the
    Hankel-based profile continues to decay.
    The inset shows the Markov lengths extracted from an exponential fit of the CMI. See SM for details on the fitting and statistical errors. When not visible, the error bars are smaller than the size of the data points.
    }
    \label{fig:comparison}
\end{figure}

To demonstrate the advantages of the Hankel truncation scheme, we numerically simulate CMI dynamics in a one-dimensional DP model and compare its performance with conventional canonical truncation at the same maximum bond dimension $\chi_{\max}=180$. As shown in Fig.~\ref{fig:comparison}, the two methods agree at short times ($t=640$). At later times ($t=1024$), however, canonical truncation develops an artificial plateau in the CMI at large buffer sizes, whereas Hankel truncation continues to accurately resolve the decaying tail. Consequently, the Markov lengths extracted from the two methods agree up to $t\simeq 900$, beyond which canonical truncation produces a spurious divergence of $\xi_M$. In contrast, Hankel truncation yields reliable estimates of $\xi_M$ up to $t\simeq 1800$. This improvement also allows us to access the asymptotic critical scaling regime, which was previously inaccessible (see End Matter).

\emph{Universal CMI scaling in PCDP ---} Armed with these numerical techniques, we now investigate the scaling of CMI in a classical stochastic model in the PCDP universality class~\cite{hinrichsen2000,odor2004,NMenyhard1994,pcdp94,pcdp26,ravindranath2023entanglement,huse25}, as illustrated in Fig.~\ref{fig:setup}(b). This is a stochastic model where the parity of the total particle number is conserved. Such dynamics is naturally realized in terms of domain walls in a one-dimensional chain of classical Ising spins $s_i=\pm 1$. 
Each domain wall between neighboring spins corresponds to a
particle, with bond occupation number
$N_{i+1/2} = \frac{1-s_is_{i+1}}{2} \in\{0,1\}$.
The allowed dynamical moves shown in Fig.~\ref{fig:setup}(b) are implemented by local spin updates
\begin{align}
    (s,s,-s)
    &\overset{P}{\longleftrightarrow}
    (s,-s,-s),
    &&(\mathrm A\emptyset
       \leftrightarrow
       \emptyset\mathrm A),
    \notag\\
    (s,-s,s)
    &\xrightarrow{R}
    (s,s,s),
    &&(2\mathrm A\to\emptyset),
    \notag\\
    (s,s,-s,-s)
    &\xrightarrow{Q}
    (s,-s,s,-s),
    &&(\mathrm A\to3\mathrm A).
    \label{eq:pcdp_updates}
\end{align}
Here $P$, $R$, and $Q$ are the update probabilities for
diffusion, pair annihilation, and branching, respectively.
The competition between branching and pair annihilation gives rise to a continuous absorbing-state phase transition. At weak branching, the system is in the inactive phase, where the long-time dynamics is dominated by diffusion-limited pair annihilation, and the domain-wall density decays as $\rho(t)\sim t^{-1/2}$~\cite{cardy1998field,odor2004}. At strong branching, the system enters an active phase with a finite domain-wall density that persists for sub-exponential times. The phase diagram as a function of the branching rate $Q$, with $P=0.50$ and $R=0.30$ fixed, is shown in Fig.~\ref{fig:setup}(c).

We initialize the spins in an unbiased product distribution
$|P_0\rangle = \bigotimes_i \left(\frac{1}{2}|\uparrow\rangle_i + \frac{1}{2}|\downarrow\rangle_i\right)$.
The stochastic evolution is organized into a four-layer
Floquet protocol with a two-site spatial unit cell \cite{SupplementalMaterial},
$\mathcal F = \mathcal B_{\rm even}(Q) \mathcal B_{\rm odd}(Q) \mathcal D_{\rm even}(P,R) \mathcal D_{\rm odd}(P,R)$.
The layers $\mathcal D_{\rm odd/even}$ simultaneously update
all odd/even sites through single-spin flips implementing
diffusion and pair annihilation. The layers $\mathcal B_{\rm odd/even}$ simultaneously update the disjoint spin pairs $(2j-1,2j)$ / $(2j,2j+1)$,
respectively, implementing branching through pair flips.
Within each layer, the stochastic updates are independent
conditioned on the configuration entering that layer~\cite{SupplementalMaterial}.
After $t$ complete periods, the probability state is
$|P_t\rangle=\mathcal F^t|P_0\rangle$. For the CMI calculations, we take $A$ and $C$ to each consist of two spins, separated by a buffer region $B$ containing $2L$ consecutive spins, corresponding to $\ell_B=2L-1$ bonds [Fig.~\ref{fig:setup}(d)].

\begin{figure*}[t]
    \centering
    \includegraphics[width=\textwidth]{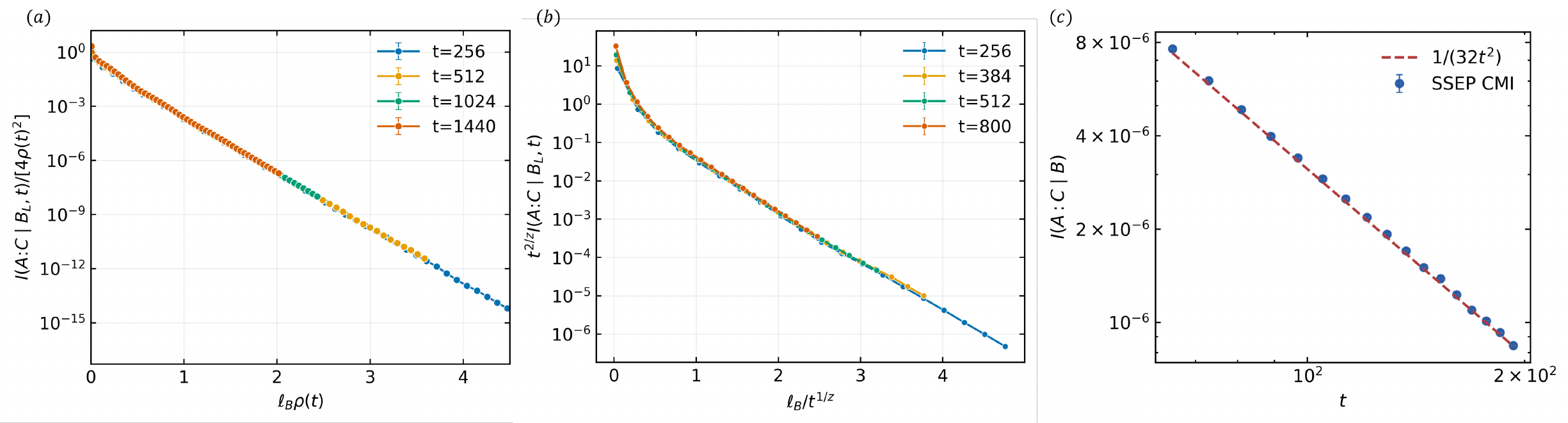}
    \caption{
    (a) Scaling collapse of CMI in the inactive phase PCDP according to Eq.~(\ref{eq:scaling_inactive}). We take $P=0.50$, $R=0.30$, and $Q=0.010$.
    Here $\ell_B=2L-1$ counts the total number of bonds
    inside a buffer. We consider buffers up to $L=150$, corresponding to
    $300$ spins and $\ell_B=299$. The domain-wall density $\rho(t)$ was computed from the same time-evolved iMPS.
    (b) Scaling collpase of CMI at the critical point $Q=0.135$ of PCDP according to Eq.~(\ref{eq:intro}) with $\ell(t)\sim t^{1/z}$, where $z=1.741$ is the dynamical exponent at PCDP critical point.
    (c) Temporal CMI versus $t$ of the infinite-system one-dimensional SSEP.
    Blue circles show the large-scale tensor-network results as a
    function of the endpoint separation $t$; the dashed line is
    the Gaussian hydrodynamic prediction
    $I(t,\infty)=1/(32t^2)$, with no additional fitting parameter.
    Error bars denote one standard sampling error and are smaller
    than the symbols where not visible.
    Numerical details are provided in the SM~\cite{SupplementalMaterial}. 
    }
    \label{fig:pcdp_collapse}
\end{figure*}

We start with the inactive phase with $Q=0.010$. In Fig.~\ref{fig:pcdp_collapse}(a), we plot the CMI in the inactive phase as a function of $\ell_B$ at different times. At each finite time, the CMI decays exponentially at large $\ell_B$, indicating a finite Markov length. Interestingly, the data for different $\ell_B$ and times collapse on top of one another when properly rescaled by the time-dependent domain-wall density $\rho(t)$:
\begin{equation}
I_t(A:C| |B|=\ell_B) = \rho(t)^2 f\big(\ell_B \rho(t) \big),
\label{eq:scaling_inactive}
\end{equation}
where the scaling function $f(x) \sim e^{-x}$ at large $x$. This scaling form immediately implies that the Markov length is set by the average domain-wall separation, $\xi_M(t)\sim \rho(t)^{-1}\sim t^{1/2}$. Thus, while the Markov length remains finite at any finite time, it diverges as $t\to\infty$ throughout the entire inactive phase. This is in stark contrast to the inactive phase in the DP process, where particles can annihilate individually, and the Markov length rapidly saturates to a time-independent value~\cite{groverdp}.

The scaling collapse in Eq.~(\ref{eq:scaling_inactive}) admits a simple physical interpretation and a straightforward derivation. In the inactive phase of PCDP, branching becomes irrelevant at long times and large length scales, and the dynamics is governed by diffusion-limited domain-wall coarsening. Intuitively, the coarsening process progressively eliminates nearby domain walls, generating spatial correlations among the surviving walls on a growing length scale set by their average separation $\ell(t)\sim\rho(t)^{-1}$, which naturally accounts for the divergence of the Markov length. More precisely, the coarsening picture suggests that
the joint statistics of wall numbers and positions become self-similar when distances are measured in units of $\ell(t)$~\cite{cardy1998field,garrod2018pfaffian}. Let $1\leq i_1<\cdots<i_k\leq n$ denote the domain wall positions in an interval of $n$ bonds and 
$p_t(i_1,\ldots,i_k;n)$ be the probability of finding walls
at exactly these positions and nowhere else.
Self-similarity implies that the joint probability can be written as
\begin{equation}
p_t(i_1,\ldots,i_k;n)
= \ell(t)^{-k}
g_k\!\left(
\frac{i_1}{\ell(t)},\ldots,\frac{i_k}{\ell(t)};
\frac{n}{\ell(t)}
\right),
\label{eq:inactive_configuration_scaling}
\end{equation}
where the function $g_k$ is time independent. As shown in the End Matter, summing over all wall numbers
and configurations yields the following result for the Shannon entropy in a region of length $n$:
\begin{equation}
H_t(n)=
-n\rho(t)\log\rho(t)
+h\big(n\rho(t)\big).
\label{eq:inactive_entropy_scaling}
\end{equation}
The first term accounts for the uncertainty of the wall positions on the
lattice, and the function $h(x)$ contains contributions from the scaling function $g_k$. Translation invariance indicates that the CMI can be written as $I_t(A:C||B|=\ell_B)= 2H_t(\ell_B+2)-H_t(\ell_B)-H_t(\ell_B+4)=2h\!\left[x+2\rho(t)\right]-h(x)
-h\!\left[x+4\rho(t)\right]$, with $x=\ell_B\rho(t)$. A simple Taylor expansion immediately yields the scaling form~(\ref{eq:scaling_inactive}), with the scaling function $f(x)=-h''(x)$.

In Fig.~\ref{fig:pcdp_collapse}(b), we compute the CMI at the critical point of the PCDP process. We find that the CMI at different times again collapses onto the scaling form~(\ref{eq:scaling_inactive}), with the characteristic length scale $\rho(t)^{-1}$ replaced by $t^{1/z}$, where $z=1.741$ is the dynamical critical exponent. The scaling function exhibits an exponential decay at large $\ell_B$, implying a Markov length that diverges as $\xi_M(t)\sim t^{1/z}$. Unlike in the inactive phase, where $\xi_M(t)$ is set by the average domain-wall separation, the Markov length at criticality is governed by the growing dynamical correlation length over which particles exhibit strong correlations. Thus, despite the distinct physical mechanisms underlying the growth of $\xi_M(t)$, the CMI obeys the same scaling form in both cases. We emphasize that the Hankel truncation scheme is essential for obtaining the scaling collapse shown in Fig.~\ref{fig:pcdp_collapse} for the PCDP process, as conventional canonical truncation becomes numerically unstable before the scaling regime is reached.

We remark that the scaling form in Eq.~(\ref{eq:intro}) extends to a broader class of classical stochastic dynamics. For instance, Ref.~\cite{hauser2026strong} investigated CMI dynamics in a one-dimensional SSEP and identified the same scaling form through a heuristic data collapse, with the Markov length growing linearly in time. Similarly, the CMI dynamics following a quench from the Ising ground state to infinite temperature obeys the same scaling form, with an exponentially growing Markov length~\cite{lloyd2025}. In both cases, the growth of the Markov length is associated with an emergent length scale governing global inference, which increases as the system evolves.

\emph{Temporal CMI in SSEP ---} 
\label{sec:temporalcmi}
To further demonstrate the generality of our proposed scaling form, we extend our analysis from spatial to temporal CMI. The definition of temporal CMI is analogous to its spatial counterpart. Specifically, consider measuring the particle occupation at a fixed site (the origin) at equally spaced times $0,1,\ldots,t$, with the system undergoing classical stochastic evolution between successive measurements. Denoting the sequence of measurement outcomes by $\{S_0(\tau)\}_{\tau=0}^t$, we define the temporal CMI as the conditional correlation between the outcomes at times $0$ and $t$, given all intermediate outcomes: $I\!\left[S_0(0):S_0(t)\,\middle|\,S_0(1),\ldots,S_0(t-1)\right]$. When this quantity decays exponentially at large $t$, the characteristic Markov time has the physical meaning of memory time. In particle-number-conserving stochastic processes such as the SSEP, slow hydrodynamic modes give rise to a Markov time that diverges with system size, as was recently pointed out in Ref.~\cite{iadecola2026local}. Explicitly, we consider the following scaling ansatz for the temporal CMI in this case:
\begin{equation}
    I(t,L)
    = \frac{1}{L^{\beta}}
    f\left(\frac{t}{L^{z}}\right),
    \label{eq:ssep_temporal_fss}
\end{equation}
where $L$ is the system size, and $t$ is the buffer region in the time direction. We now argue that the exponent $\beta=2z$, so that the scaling ansatz~(\ref{eq:ssep_temporal_fss}) is nothing but Eq.~(\ref{eq:intro}) in the time domain.

Let us consider the CMI in the thermodynamic limit $L\rightarrow \infty$. In this case, we expect that the CMI in Eq.~(\ref{eq:ssep_temporal_fss}) becomes independent of $L$, which requires $f(x)\sim x^{-\alpha}$ as $x\rightarrow 0$. Matching with the scaling ansatz~(\ref{eq:ssep_temporal_fss}), we conclude that the infinite-size CMI decays as a power-law in time $I(t,\infty)\sim 1/t^{\alpha}$, and $\beta=\alpha z$. We derive in the End Matter using Gaussian fluctuating hydrodynamics that
\begin{equation}
    I(t,\infty)=\frac{1}{32t^2},
\end{equation}
which immediately fixes $\alpha=2$ and $\beta=2z$. To test the validity of Gaussian hydrodynamic prediction, we numerically compute the CMI of the infinite one-dimensional SSEP by contracting the (1+1)-dimensional spacetime tensor network along the spatial direction. Remarkably, as shown in Fig.~\ref{fig:pcdp_collapse}(c), the numerical results agree quantitatively with the Gaussian hydrodynamic prediction, including the exact prefactor $1/32$. We note that Ref.~\cite{iadecola2026local} previously estimated $\beta\approx 3.2(1)$ from scaling collapses of relatively small systems. Our analytical and numerical results suggest that this deviation from the predicted value $\beta=4$ ($z=2$ from diffusive transport) arises from finite-size effects.

\emph{Discussion}---
We propose a simple universal scaling form for CMI dynamics in classical stochastic processes with a diverging Markov length. Our explicit derivation in the inactive phase of a one-dimensional PCDP model makes the physical origin of this scaling transparent: the inverse-square prefactor follows directly from translation invariance and the fixed sizes of the endpoint regions, making it robust across a broad class of stochastic dynamics with different growth laws for the Markov length. We further extend the scaling analysis to temporal CMI, where slow hydrodynamic modes give rise to a diverging Markov time. 
To reliably resolve the small, decaying tails of CMI, we develop a novel Hankel-based truncation scheme tailored to iMPS representations of classical probability distributions, enabling simulations at significantly longer times and larger system sizes. Given the widespread use of MPS-based simulations of classical master equations in studies of operator scrambling, nonequilibrium stochastic dynamics, and related problems, our numerical method may also be of independent interest. It will be an interesting open question to further explore the validity and possible mechanism for its breakdown in other dynamical settings, including classical stochastic dynamics with distinct transport universality classes, systems without translation invariance, higher dimensional geometry, as well as genuinely quantum dynamics.

\emph{Acknowledgments.---}
This work is supported by Grant No. 12375027 from the National Natural Science Foundation of China. Numerical simulations were performed on the High-performance Computing Platform of Peking University.

\let\savedaddcontentsline\addcontentsline
\renewcommand{\addcontentsline}[3]{}
\bibliography{references}
\let\addcontentsline\savedaddcontentsline

\clearpage

\onecolumngrid
\begin{center}
\bfseries End Matter
\end{center}
\vspace{0.6em}

\twocolumngrid

\emph{Probability iMPS: environment and observables ---}
A translation-invariant probability iMPS assigns a matrix
$A^s$ to each local physical degree of freedom $s$.
Marginalizing an unobserved site sums over all of its possible
outcomes and therefore replaces its local tensor by the
transfer matrix
$T=\sum_s A^s$. To see how the environment enters, consider a long but finite
open chain whose matrix-product contraction is closed by
boundary vectors $\langle e_{\rm L}|$ and
$|e_{\rm R}\rangle$.
These vectors encode the boundary conditions at the two
physical ends of the chain.
Let a sequence of observed sites $w=(s_1,\ldots,s_n)$ be separated from
the left and right ends by $N_{\rm L}$ and $N_{\rm R}$
unobserved sites, respectively.
Writing $A^w=A^{s_1}\cdots A^{s_n}$, marginalizing the
unobserved sites gives the normalized probability
\begin{equation}
    p_{N_{\rm L},N_{\rm R}}(w)
    =
    \frac{
    \langle e_{\rm L}|
    T^{N_{\rm L}}A^wT^{N_{\rm R}}
    |e_{\rm R}\rangle
    }{
    \langle e_{\rm L}|
    T^{N_{\rm L}+n+N_{\rm R}}
    |e_{\rm R}\rangle
    }.
    \label{eq:em_finite_marginal}
\end{equation}
Assuming $T$ has a unique dominant eigenvalue
$\lambda_0$, repeated applications of $T$ suppress all
subleading modes relative to the dominant one.
In the thermodynamic limit, the two long unobserved regions
in Eq.~\eqref{eq:em_finite_marginal} are therefore represented
by the dominant left and right eigenvectors,
$\langle\mathcal L|$ and $|\mathcal R\rangle$.
We rescale each $A^s$ so that $\lambda_0=1$, giving the
fixed-point relations
\begin{equation}
    \langle\mathcal L|T=\langle\mathcal L|,
    \qquad
    T|\mathcal R\rangle=|\mathcal R\rangle,
    \label{eq:em_transfer_fixed_points}
\end{equation}
and further impose the normalization
$\langle\mathcal L|\mathcal R\rangle=1$.
In the thermodynamic limit, the finite-chain boundary factors
then cancel between the numerator and denominator of
Eq.~\eqref{eq:em_finite_marginal}, leaving $p(w) = \langle\mathcal L|A^w|\mathcal R\rangle$.

\begin{figure}[t]
    \centering
    \includegraphics[width=\columnwidth]{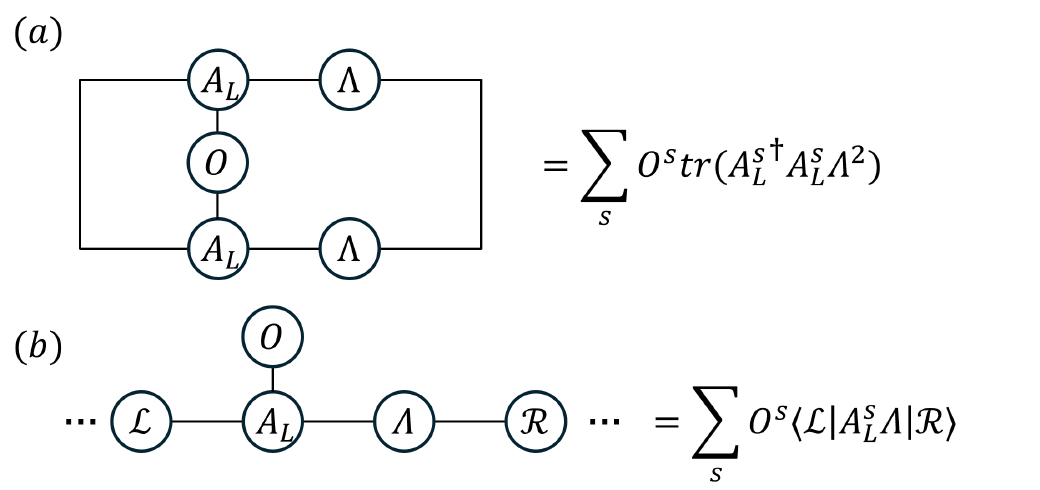}
    \caption{
    Schmidt weights enter observables in quantum and probability MPS differently.
    (a) A diagonal one-site observable in a quantum MPS is
    evaluated through a double-layer contraction, which
    weights the virtual sectors through $\Lambda^2$ and aligns
    their contributions with the Schmidt ordering.
    (b) The corresponding classical observable in a
    probability iMPS is evaluated through a single-layer
    contraction.
    The bond spectrum then enters linearly together with the
    left and right environments, so a canonical weight alone
    does not determine the importance of the associated
    virtual direction.
    }
    \label{fig:contractions}
\end{figure}

To see how the boundary environments affect probability
measurements, consider a one-site observable.
For a quantum MPS, Schmidt ordering is not only optimal in a state
vector 2-norm, but also aligned with the way local observables are evaluated.
In the canonical gauge, as illustrated schematically in Fig.~\ref{fig:contractions}(a), a diagonal one-site observable has the form
\begin{equation}
    \langle\widehat O\rangle_\psi =
    \sum_s O^s\,
    \operatorname{tr}\!\left[
        (A_{\rm L}^{s})^\dagger
        A_{\rm L}^{s}\Lambda^2
    \right],
    \label{eq:em_quantum_double_layer}
\end{equation}
where $\Lambda$ contains the Schmidt coefficients and
$A_{\rm L}^s$ are the left-canonical tensors satisfying $\sum_s (A_{\rm L}^s)^\dagger A_{\rm L}^s = I_{\chi}$.
The two layers of the quantum contraction therefore weigh
the virtual sectors through $\Lambda^2$, aligning their
contributions to the observable with the Schmidt ordering.

For a probability iMPS, the same observable is evaluated by a
single-layer contraction.
Writing an analogous canonical-like factorization, we obtain
\begin{align}
    \langle O\rangle_p
    &=
    \sum_s O^s
    \langle\mathcal L|
    A_{\rm L}^{s}\Lambda
    |\mathcal R\rangle .
    \label{eq:em_probability_single_layer}
\end{align}
In the single-layer contraction, the Schmidt coefficients in
$\Lambda$ enter only linearly and together with the left and
right environments.
They therefore do not by themselves rank virtual directions
according to their contributions to probability contractions.
This mismatch is especially consequential for a strongly
nonnormal transfer matrix, where the boundary
fixed-point problem becomes ill conditioned: the resulting
environments are highly sensitive to repeated evolution and
truncation and can amplify directions associated with small
Schmidt coefficients. 

\paragraph{Interval entropy from self-similar coarsening ---}
We derive the interval-entropy scaling used in
Eq.~\eqref{eq:inactive_entropy_scaling} from the
configuration-level ansatz in
Eq.~\eqref{eq:inactive_configuration_scaling}.
Let $\boldsymbol i=(i_1,\ldots,i_k)$, with
$1\leq i_1<\cdots<i_k\leq n$, denote a configuration
containing exactly $k$ walls in an interval of $n$ bonds.
Its entropy is
\begin{equation}
    H_t(n)
    =
    -\sum_{k=0}^{n}
    \sum_{i_1<\cdots<i_k}
    p_t(\boldsymbol i;n)
    \log p_t(\boldsymbol i;n),
    \label{eq:em_wall_interval_entropy}
\end{equation}
where $k=0$ denotes the empty configuration.

Substituting
$p_t(\boldsymbol i;n) =
\ell(t)^{-k}g_k[\boldsymbol i/\ell(t);n/\ell(t)]$
gives
\begin{align}
    H_t(n)
    &=
    \langle K\rangle_t\log\ell(t)
    \notag\\
    &\quad-
    \sum_{k=0}^{n}
    \sum_{i_1<\cdots<i_k}
    \ell(t)^{-k}g_k
    \log g_k,
    \label{eq:em_entropy_decomposition}
\end{align}
where $K$ is the number of walls in the interval.
At the coarse-grained level,
$\langle K\rangle_t\simeq n\rho(t)$.

In the scaling limit $\ell(t)\to\infty$ at fixed
$x=n/\ell(t)$, the sums over lattice positions become
integrals over the rescaled wall coordinates
$y_j=i_j/\ell(t)$.
We therefore obtain
\begin{align}
    H_t(n)
    &\simeq
     \langle K\rangle_t\log\ell(t)
    +
    h\!\left(\frac{n}{\ell(t)}\right),
    \notag\\
    h(x)
    &=
    -\sum_{k=0}^{\infty}
    \int_{0<y_1<\cdots<y_k<x}
    g_k(\boldsymbol y;x)
    \log g_k(\boldsymbol y;x)\,d^ky.
    \label{eq:em_inactive_entropy_scaling}
\end{align}
For $k=0$, the integral is understood as
$-g_0(x)\log g_0(x)$.

\paragraph{Gaussian fluctuating hydrodynamics prediction for temporal CMI ---}
For convenience, we subtract from the local densities its average value at half-filling
$X_j=S_0(j)-1/2$.
To describe the long-time hydrodynamic regime, we approximate
the stationary joint distribution of
$(X_0,\ldots,X_t)$ by a multivariate Gaussian.
Linear fluctuating hydrodynamics in one dimension then gives
the corresponding two-time covariance as~\cite{bertini2015mft}
\begin{equation}
    C_k
    \equiv
    \langle X_jX_{j+k}\rangle
    \simeq
    \frac{\chi}{\sqrt{4\pi D|k|}}
    \equiv
    c\,|k|^{-1/2},
    \qquad |k|\gg1.
    \label{eq:em_temporal_covariance}
\end{equation}
In the space--time units used for our Floquet protocol, the
effective diffusion constant is $D=1$, while half filling gives
$\chi=\operatorname{Var}(S_i)=1/4$.

Due to stationarity in the long-time regime, the temporal CMI can be similarly expressed in terms of an interval entropy in time domain $I_{\rm G}(t)=2H_t-H_{t-1}-H_{t+1}$,
where $H_t$ is the entropy of
$(X_0,\ldots,X_{t-1})$.
For Gaussian probabilities, the block entropy can be readily evaluated as
\begin{equation}
    H_t = \frac{1}{2}
    \log\!\left[(2\pi e)^t\det\Gamma_t\right],
    \label{eq:em_gaussian_block_entropy}
\end{equation}
where $\Gamma_t$ is the covariance matrix and stationarity
makes it Toeplitz,
\begin{equation}
    (\Gamma_m)_{jk} = \langle X_jX_k\rangle
    = C_{j-k}, \qquad 0\leq j,k\leq t-1.
    \label{eq:em_temporal_toeplitz}
\end{equation}

To determine the large-$t$ behavior of
$\det\Gamma_t$, we introduce the power spectral density
\begin{equation}
    f(\omega)
    =
    \sum_{k\in\mathbb Z}C_k e^{-ik\omega},
    \label{eq:em_temporal_spectral_density}
\end{equation}
whose Fourier coefficients are the stationary covariances
$C_k$.
The hydrodynamic tail
$C_k\sim c|k|^{-1/2}$ produces a low-frequency singularity
$f(\omega)\sim c'|\omega|^{-1/2}$.
Since $|1-e^{i\omega}|\sim|\omega|$ near $\omega=0$, we
separate this singularity by writing
\begin{equation}
    f(\omega)
    =
    |1-e^{i\omega}|^{2\alpha}g(\omega),
\end{equation}
where $g(\omega)$ contains the regular part of the spectrum.
Matching the power law gives $2\alpha=-1/2$, and hence
$\alpha=-1/4$.
The refined Fisher--Hartwig expansion yields
~\cite{deift2011fisherhartwig}
\begin{equation}
    \log\det\Gamma_t
    =
    a_0t+\alpha^2\log t+a_1
    +\frac{a_2}{t}
    +O(t^{-2}),
    \label{eq:em_fisher_hartwig_determinant}
\end{equation}
where the coefficients $a_j$ are independent of $m$ and are
determined by the regular background $g(\omega)$ together
with the singularity exponent $\alpha$.

Combining Eq.~\eqref{eq:em_fisher_hartwig_determinant} with
the Gaussian entropy in
Eq.~\eqref{eq:em_gaussian_block_entropy}, we separate the
extensive and subextensive contributions as
\begin{equation}
    H_t
    =
    h_0t+\delta H_t,
    \qquad
    \delta H_t
    =
    \frac{\alpha^2}{2}\log t+h_1
    +\frac{h_2}{t}
    +O(t^{-2}),
    \label{eq:em_gaussian_entropy_asymptotics}
\end{equation}
which yields
\begin{align}
    I_{\rm G}(t)
    &=
    \frac{\alpha^2}{2}
    \log\!\left[
        \frac{t^2}{(t-1)(t+1)}
    \right]
    +O(t^{-3})
    \notag\\
    &=
    \frac{1}{32t^2}
    \left[1+O(t^{-1})\right].
    \label{eq:em_gaussian_cmi_asymptotics}
\end{align}

\begin{figure}[t]
    \centering
    \includegraphics[width=0.85\columnwidth]{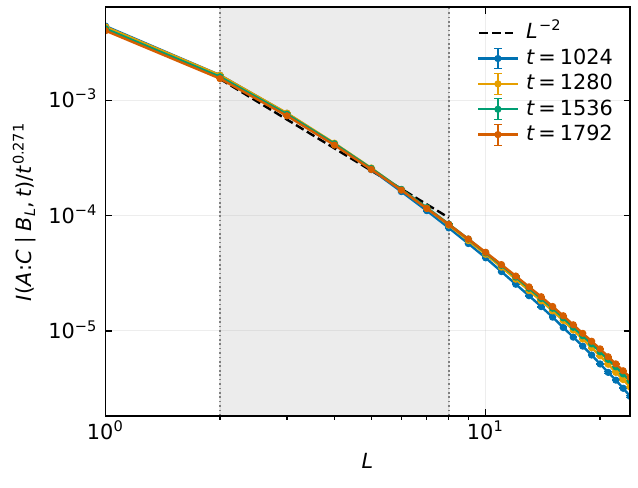}
    \caption{ CMI scaling of DP at the critical point in the regime where $1 \lesssim \ell_B \ll \ell(t)\sim t^{1/z}$. We find a window where CMI follows a power-law behavior $\sim 1/\ell_B^2$.
    The effective time exponent $0.271$ is obtained by
    a least-squares fit over $2\leq \ell_B\leq8$ (shaded region)
    and $1024\leq t\leq1792$.
    Error bars denote one sampling standard error and are
    smaller than the symbols.
    }
    \label{fig:dp_smallL_collapse}
\end{figure}

\paragraph{CMI of DP in the critical scaling regime---}
In Fig.~\ref{fig:dp_smallL_collapse}, we show a scaling collapse of the CMI at the critical point of DP in the regime $1\lesssim \ell_B\ll \ell(t)$, where $\ell(t)\sim t^{1/z}$ is the dynamically generated spatial correlation length. Within the limited accessible window, the CMI approximately follows a power law, $I_t\sim \ell_B^{-2}$. Our numerical fits also suggest a weak time dependence, $I_t\sim t^{0.271}$ at fixed $\ell_B$, although the limited fitting window prevents us from drawing a conclusion about this exponent. Nevertheless, this late-time critical scaling regime was previously inaccessible to iMPS simulations using conventional canonical truncation.

\clearpage
\onecolumngrid

\setcounter{section}{0}
\setcounter{subsection}{0}
\setcounter{equation}{0}
\setcounter{figure}{0}
\setcounter{table}{0}
\setcounter{algorithm}{0}
\setcounter{secnumdepth}{2}

%%begin novalidate
\makeatletter
\providecommand{\href@noop}[2]{#2}
\makeatother
%%end novalidate

\renewcommand{\thesection}{S\arabic{section}}
\renewcommand{\thesubsection}{\thesection.\Alph{subsection}}
\renewcommand{\theequation}{S\arabic{equation}}
\renewcommand{\thefigure}{S\arabic{figure}}
\renewcommand{\thetable}{S\arabic{table}}
\renewcommand{\thealgorithm}{S\arabic{algorithm}}

\makeatletter
\renewcommand{\p@subsection}{}
\makeatother

\renewcommand{\theHsection}{supp.\arabic{section}}
\providecommand{\theHsubsection}{}
\renewcommand{\theHsubsection}{supp.\arabic{section}.\Alph{subsection}}
\renewcommand{\theHequation}{supp.\arabic{equation}}
\renewcommand{\theHfigure}{supp.\arabic{figure}}
\renewcommand{\theHtable}{supp.\arabic{table}}
\providecommand{\theHalgorithm}{}
\renewcommand{\theHalgorithm}{supp.\arabic{algorithm}}
\begin{center}
{\large\bf Supplemental Material for\\[2pt]
Universal Scaling of Conditional Mutual Information in Classical Stochastic Dynamics}
\end{center}

% List supplemental sections and subsections.
\addtocontents{toc}{\protect\setcounter{tocdepth}{2}}
\setcounter{tocdepth}{2}
\tableofcontents

\section{Probability iMPS and the sampled CMI estimator}
\label{sec:sampling}

\subsection{Sequential sampling and the CMI estimator}
The CMI can then be evaluated directly from these finite-region
marginals. For three adjacent regions $A$, $B$, and $C$, it takes
the form
\begin{equation}
    I(A:C\mid B)
    =\sum_{b} p(b)
    \sum_{a,c} p(a,c\mid b)
    \ln\frac{p(a,c\mid b)}
    {p(a\mid b)p(c\mid b)},
    \label{eq:cmi}
\end{equation}
where $a$, $b$, and $c$ denote configurations of the corresponding
regions. Thus, the finite-region probabilities encoded by the iMPS
are the basic inputs needed to evaluate CMI. Rather than enumerate the exponentially many buffer
configurations, we evaluate the buffer average of Eq.(\ref{eq:cmi}) by drawing $N_s$ independent samples
$\hat b^{(n)}\sim p(b)$ from the distribution represented by
the iMPS~\cite{lloyd2025}.
The resulting estimator is
\begin{equation}
\widehat I(A:C\mid B_L)
=
\frac{1}{N_s}
\sum_{n=1}^{N_s}
I(A:C\mid B_L=\hat b^{(n)}).
\label{eq:sestimator}
\end{equation}
For exact sampling from the represented distribution, this
estimator is unbiased.

To generate buffer configurations $p(b)$, we use the probability chain rule:
\begin{equation}
    p(s_1,\ldots,s_L) =\prod_{k=1}^{L}p(s_k\mid s_{1:k-1}),
    \label{eq:schainrule}
\end{equation}
where $s_{1:k-1}=(s_1,\ldots,s_{k-1})$ and $s_{1:0}$ is the
empty prefix.
Before drawing the $k$th symbol, the prefix
$\hat s_{1:k-1}$ has already been sampled.
We encode its effect on the remaining chain in the normalized
left boundary message
\begin{equation}
    \langle m_{k-1}| =
    \frac{\langle\mathcal L| A^{\hat s_1}\cdots A^{\hat s_{k-1}}}{ p(\hat s_{1:k-1})}, \qquad \langle m_0|=\langle\mathcal L|.
    \label{eq:sleftmessage}
\end{equation}
For each possible next symbol $s$, this message gives the
conditional probability
\begin{align}
    q_k(s) &\equiv p(s_k=s\mid\hat s_{1:k-1}) =
    \langle m_{k-1}|A^s|\mathcal R\rangle,
    \notag\\ \langle m_k| &=
    \frac{\langle m_{k-1}|A^{\hat s_k}}{q_k(\hat s_k)}.
    \label{eq:ssamplingrecursion}
\end{align}
We draw $\hat s_k$ according to $q_k$ and then apply the
second line to update the boundary message.
Since $T=\sum_s A^s$ and
$T|\mathcal R\rangle=|\mathcal R\rangle$, these probabilities
are normalized:
\begin{equation}
\sum_s q_k(s) = \langle m_{k-1}|T|\mathcal R\rangle = 1.
\end{equation}
Repeating this recursion samples the buffer from the
probability distribution represented by the iMPS.

\subsection{Conditional endpoint distributions}
To compute the conditional endpoint distribution, we also retain
the left-endpoint value $a$ in a separate boundary message,
\begin{equation}
\langle u_{a,k}|
=
\frac{\langle\mathcal L|A^a
A^{\hat s_1}\cdots A^{\hat s_k}}
{p(\hat s_{1:k})},
\qquad
\langle u_{a,0}|=\langle\mathcal L|A^a.
\label{eq:sendpointmessage}
\end{equation}
For every sampled symbol, all endpoint messages are updated
alongside $\langle m_k|$ using
\begin{equation}
\langle u_{a,k}|
=
\frac{\langle u_{a,k-1}|A^{\hat s_k}}
{q_k(\hat s_k)}.
\label{eq:sendpointupdate}
\end{equation}
Contracting with the right-endpoint tensor then gives
\begin{equation}
p(a,c\mid\hat s_{1:k})
=
\frac{\langle\mathcal L|A^a
A^{\hat s_1}\cdots A^{\hat s_k}A^c
|\mathcal R\rangle}
{p(\hat s_{1:k})}
=
\langle u_{a,k}|A^c|\mathcal R\rangle.
\label{eq:sendpointtable}
\end{equation}
At each step $k$, the propagated messages yield the full
conditional endpoint distribution
$p(a,c\mid\hat s_{1:k})$.
Its row and column sums give the corresponding conditional
marginals, from which we evaluate the record-resolved mutual
information
$I(A:C\mid B_k=\hat s_{1:k})$.
We generate each sampled buffer up to a maximum length
$L_{\max}$.
For any $k\leq L_{\max}$, its first $k$ symbols form a sample
from the marginal distribution of the length-$k$ buffer
$B_k$.
The same trajectory therefore yields one record-resolved
mutual information for every buffer length
$k=1,\ldots,L_{\max}$.
At each fixed $k$, averaging these values over the
$N_s$ independent trajectories gives the estimator in
Eq.~\eqref{eq:sestimator}.
Because the estimates at different $k$ are obtained from
prefixes of the same trajectories, they are statistically
correlated.
Our uncertainty analysis preserves these correlations by
resampling complete CMI profiles, as described in
Sec.~\ref{sec:numerical_details}.

\subsection{Vector-message implementation and cost}
To understand the computational advantage of this implementation,
we compare it with the full-matrix formulation~\cite{lloyd2025}, which propagates the normalized string operator
\begin{equation}
F_k=
\frac{A^{\hat s_1}\cdots A^{\hat s_k}}
     {p(\hat s_1,\ldots,\hat s_k)}
=\frac{F_{k-1}A^{\hat s_k}}{q_k(\hat s_k)},
\qquad F_0= I.
\label{eq:sfullmessage}
\end{equation}
The boundary messages used in our implementation are obtained
by contracting $F_k$ with $\langle\mathcal L|$ and
$\langle\mathcal L|A^a$:
\begin{equation}
\langle m_k|=\langle\mathcal L|F_k,
\qquad
\langle u_{a,k}|=\langle\mathcal L|A^aF_k.
\label{eq:svectoractions}
\end{equation}
Our recursion updates these messages directly, without constructing
the full matrix $F_k$.

The message recursion yields the same sampling probabilities
and conditional endpoint distributions as the full-matrix
formulation, using only matrix--vector contractions.
At fixed local dimension $d$, computing the CMI curve up to
$L_{\max}$ from $N_s$ trajectories therefore costs
$O(N_sL_{\max}\chi^2)$, compared with
$O(N_sL_{\max}\chi^3)$ for the full-matrix recursion
\cite{lloyd2025}.
This reduces the contraction cost without changing the sampling
variance at fixed $N_s$; resolving weak CMI may still require
many trajectories.

\section{Length-discounted Hankel truncation}
\label{sec:Hankel_balanced_truncation}
We now describe the probability-aware truncation method used
in our simulations.
Before truncation, let the probability iMPS have bond
dimension $D$ and local matrices
$A^s\in\mathbb R^{D\times D}$.
We seek a reduced representation
$A_\chi^s\in\mathbb R^{\chi\times\chi}$, designed to preserve the finite-word
probabilities relevant to the CMI calculation.

\subsection{Discounted Hankel matrix and virtual-space Gramians}
We introduce the length-discounted Hankel matrix and absorb
the discount into the local tensors by defining
$A_\beta^s=\sqrt{\beta}\,A^s$:
\begin{equation}
    \begin{aligned}
        H_\beta(u,v) & \equiv
        \beta^{(|u|+|v|)/2}p(uv)\\
        &=\langle\mathcal L|
        A_\beta^uA_\beta^v
        |\mathcal R\rangle,
        \qquad 0<\beta<1.
    \end{aligned}
    \label{eq:shankel}
\end{equation}
Here $u$ and $v$ denote finite prefix and suffix words,
respectively.
The discount serves two purposes.
First, it exponentially downweights finite-region
probabilities with lengths much larger than
$\ell_\beta\sim(1-\beta)^{-1}$, thereby selecting the range of
region sizes emphasized by the truncation.
Second, it regularizes the infinite Hankel matrix and makes its
singular-value decomposition well defined.
Specifically, for a normalized nonnegative distribution $\sum_{|w|=n}p(w)=1$ and hence $\sum_{|w|=n}p(w)^2\leq1$. Each word of length $n$ admits $n+1$ prefix--suffix decompositions, so
\begin{align}
    \lVert H_\beta\rVert_{\rm HS}^2
    &=
    \sum_{u,v}
    \beta^{|u|+|v|}
    |p(uv)|^2
    \notag\\
    &=
    \sum_{n=0}^{\infty}
    (n+1)\beta^n
    \sum_{|w|=n}p(w)^2
    \notag\\
    &\leq
    \sum_{n=0}^{\infty}(n+1)\beta^n
    =
    \frac{1}{(1-\beta)^2}.
    \label{eq:shankel_hs}
\end{align}
Thus, $H_\beta$ is Hilbert--Schmidt and therefore compact. It consequently admits a singular-value decomposition~\cite{balle2019}.

\begin{figure}[t]
    \centering
    \includegraphics[width=0.92\columnwidth]{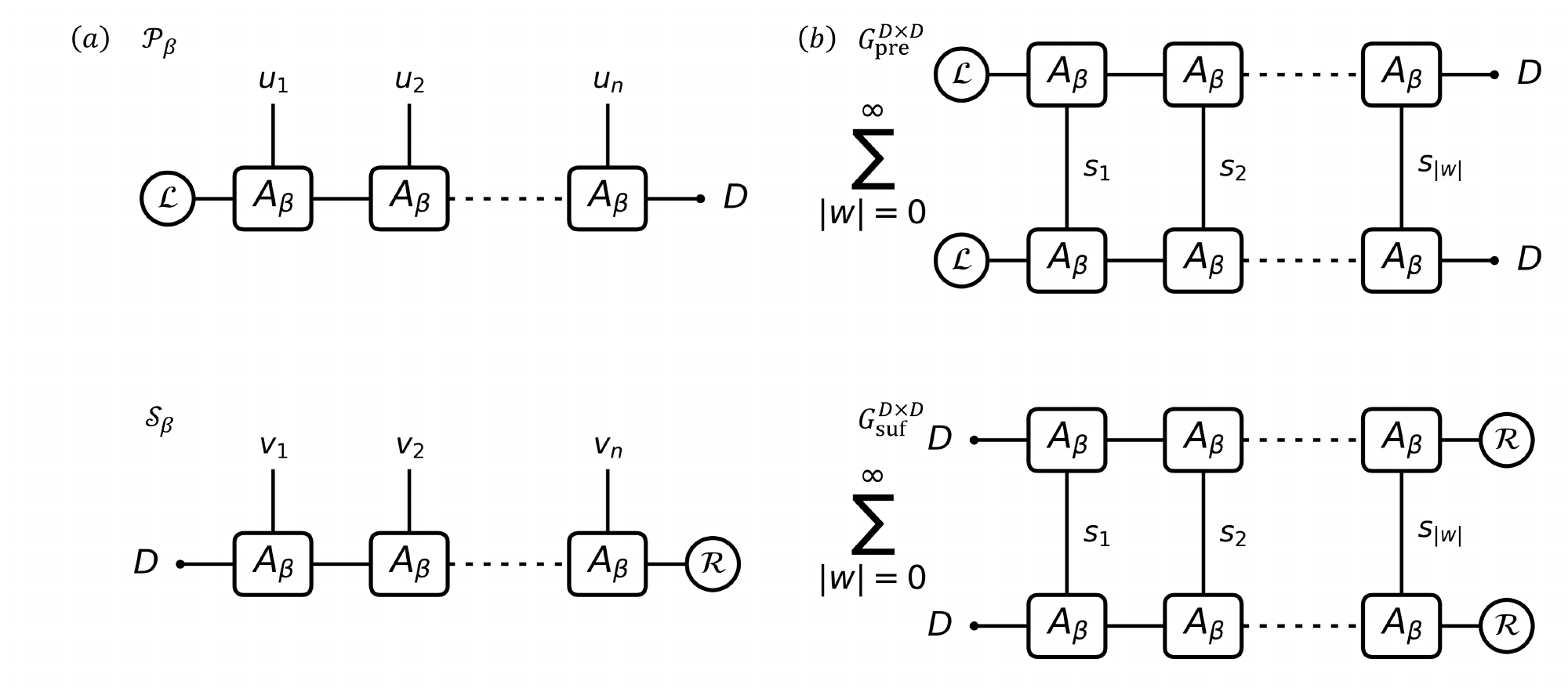}
    \caption{
    Tensor-network representation of the discounted prefix and suffix
    factors and their virtual-space Gramians.
    The upper-left and lower-left diagrams represent a row of
    $\mathcal P_\beta$ and a column of $\mathcal S_\beta$,
    respectively, for fixed words $u=(u_1,\ldots,u_n)$ and
    $v=(v_1,\ldots,v_n)$.
    The right-hand diagrams show the corresponding contractions
    $G_{\rm pre}=\mathcal P_\beta^T\mathcal P_\beta$ and
    $G_{\rm suf}=\mathcal S_\beta\mathcal S_\beta^T$.
    Each rounded square denotes
    $A_\beta^s=\sqrt{\beta}\,A^s$; vertical bonds contract the same
    physical index in the two layers, and the open virtual legs have
    dimension $D$.
    The sum includes all word lengths, with the $|w|=0$ term given by
    the corresponding boundary-vector outer product.
    }
    \label{fig:shankel_factors_gramians}
\end{figure}

To obtain the singular values without explicitly constructing
$H_\beta$, we write $H_\beta$ in a matrix form as
\begin{equation}
    H_\beta=\mathcal P_\beta\mathcal S_\beta,
    \label{eq:shankel_factorization}
\end{equation}
Each row of $\mathcal P_\beta$ is a discounted prefix vector,
and each column of $\mathcal S_\beta$ is a discounted suffix
vector, as illustrated in Fig.~\ref{fig:shankel_factors_gramians}(a)
~\cite{balle2014forward,balle2019}:
\begin{equation}
\mathcal P_\beta(u,:)
=\langle\mathcal L|A_\beta^u,
\qquad
\mathcal S_\beta(:,v)
=A_\beta^v|\mathcal R\rangle.
\label{eq:sprefix_suffix_factors}
\end{equation}
We then introduce two $D\times D$ Gramians acting on
the virtual space:
\begin{align}
G_{\rm pre}
&=\mathcal P_\beta^T\mathcal P_\beta
=\sum_u (A_\beta^u)^T
|\mathcal L\rangle\langle\mathcal L|A_\beta^u,
\notag\\
G_{\rm suf}
&=\mathcal S_\beta\mathcal S_\beta^T
=\sum_v A_\beta^v
|\mathcal R\rangle\langle\mathcal R|(A_\beta^v)^T.
\label{eq:sgramiansums}
\end{align}
These Gramians quantify the contributions of virtual
directions to the prefix and suffix factors, respectively~\cite{moore1981,antoulas2005,
benner2011,petreczky2013,balle2019}.
Separating the empty-word contribution yields the closed
linear matrix equations
\begin{align}
G_{\rm pre}
-\sum_s(A_\beta^s)^TG_{\rm pre}A_\beta^s
&=|\mathcal L\rangle\langle\mathcal L|,
\notag\\
G_{\rm suf}
-\sum_sA_\beta^sG_{\rm suf}(A_\beta^s)^T
&=|\mathcal R\rangle\langle\mathcal R|.
\label{eq:sstein}
\end{align}
which can be solved without enumerating words; in practice
we solve them iteratively and monitor their relative
residuals.

\subsection{Finite-dimensional SVD and balanced truncation}
Let
$r_{\rm pre}=\operatorname{rank}(G_{\rm pre})$ and
$r_{\rm suf}=\operatorname{rank}(G_{\rm suf})$.
Factoring the Gramians on their nonzero supports gives
\begin{equation}
G_{\rm pre}=R_{\rm pre}R_{\rm pre}^T,
\qquad
G_{\rm suf}=R_{\rm suf}R_{\rm suf}^T,
\end{equation}
where
$R_{\rm pre}\in\mathbb R^{D\times r_{\rm pre}}$ and
$R_{\rm suf}\in\mathbb R^{D\times r_{\rm suf}}$.
Since
$G_{\rm pre}=\mathcal P_\beta^T\mathcal P_\beta$ and
$G_{\rm suf}=\mathcal S_\beta\mathcal S_\beta^T$,
the prefix and suffix factors admit the support-restricted
factorizations
\begin{equation}
\mathcal P_\beta
=\mathcal Q_{\rm pre}R_{\rm pre}^T,
\qquad
\mathcal S_\beta
=R_{\rm suf}\mathcal Q_{\rm suf}^T,
\label{eq:sprefix_suffix_isometries}
\end{equation}
where $\mathcal Q_{\rm pre}^T\mathcal Q_{\rm pre} = I_{\rm pre}$ and $\mathcal Q_{\rm suf}^T\mathcal Q_{\rm suf}=I_{\rm suf}$.

Substituting Eq.~\eqref{eq:sprefix_suffix_isometries} into
Eq.~\eqref{eq:shankel_factorization} gives
\begin{equation}
    H_\beta
    =
    \mathcal Q_{\rm pre}
    K
    \mathcal Q_{\rm suf}^T,
    \qquad
    K=R_{\rm pre}^TR_{\rm suf}.
    \label{eq:shankel_core}
\end{equation}
Since $\mathcal Q_{\rm pre}$ and $\mathcal Q_{\rm suf}$ have
orthonormal columns, multiplication by these outer factors
does not change the nonzero singular values.
Consequently, the infinite-dimensional Hankel matrix
$H_\beta$ and the finite
$r_{\rm pre}\times r_{\rm suf}$ core matrix $K$ have the
same nonzero singular spectrum.
We therefore obtain the Hankel singular values from the
compact singular-value decomposition $ K=U\Sigma V^T $~\cite{balle2019}.

\paragraph{Truncation maps.---}
Let the diagonal entries of $\Sigma$ be ordered as
$\sigma_1\geq\sigma_2\geq\cdots>0$.
We retain the $\chi$ largest values and denote the
corresponding blocks by $U_\chi \in \mathbb R^{r_{\rm pre} \times \chi}$, $V_\chi \in \mathbb R^{r_{\rm suf} \times \chi}$, and
$\Sigma_\chi=\operatorname{diag}
(\sigma_1,\ldots,\sigma_\chi)$.
The associated truncation maps,
$B_\chi\in\mathbb R^{D\times\chi}$ and
$F_\chi\in\mathbb R^{\chi\times D}$, are
\begin{equation}
    B_\chi
    =
    R_{\rm suf}V_\chi\Sigma_\chi^{-1/2},
    \qquad
    F_\chi
    =
    \Sigma_\chi^{-1/2}
    U_\chi^T R_{\rm pre}^T,
    \label{eq:struncationmaps}
\end{equation}
and satisfy $F_\chi B_\chi=I_\chi$.

By construction, these maps balance the prefix and suffix
Gramians on the retained subspace~\cite{moore1981,antoulas2005,balle2019}:
\begin{equation}
    B_\chi^T G_{\rm pre}B_\chi
    =
    F_\chi G_{\rm suf}F_\chi^T
    =
    \Sigma_\chi.
    \label{eq:sbalancedgramians}
\end{equation}
Hence, each retained direction $\alpha$ has equal discounted
strength $\sigma_\alpha$ in the prefix and suffix contractions.
Keeping the largest $\sigma_\alpha$ therefore preserves the
directions with the largest contributions to the length-discounted
Hankel matrix.

Having selected the retained virtual subspace, we apply the
truncation maps to the original, undiscounted local tensors:
\begin{equation}
    A_\chi^s
    =
    F_\chi A^s B_\chi.
    \label{eq:sprojected_representation}
\end{equation}
We then form the reduced transfer matrix
$T_\chi=\sum_s A_\chi^s$ and rescale the tensors so that its
dominant eigenvalue is unity.
The corresponding left and right fixed points,
$\langle\mathcal L_\chi|$ and
$|\mathcal R_\chi\rangle$, are recomputed and normalized
according to
$\langle\mathcal L_\chi|\mathcal R_\chi\rangle=1$.
These steps restore normalization and marginal consistency
of the reduced matrix-product representation.
Implementation details are provided in Sec.~\ref{sec:numerical_details}.

\section{Numerical workflow and statistical inference}
\label{sec:numerical_details}
This section combines the CMI sampling method of
Sec.~\ref{sec:sampling} with the Hankel-based truncation
described in Sec.~\ref{sec:Hankel_balanced_truncation} to define the
numerical workflow for probability-iMPS evolution and CMI measurement.
Algorithm~\ref{alg:evolution_sampling} summarizes this
workflow, followed by the numerical consistency checks used
during the evolution.
Algorithm~\ref{alg:mkl_inference} describes the estimation of
sampling uncertainties and the extraction of Markov lengths
through block-bootstrap refitting.
Model-specific implementations and numerical settings are
given in Secs.~\ref{sec:dp_benchmark} and
\ref{sec:pcdp_protocol}.

\begin{algorithm}[H]
\caption{Probability-iMPS evolution and CMI measurement.}
\label{alg:evolution_sampling}
\small
\begin{algorithmic}[1]
\Statex \textbf{Input:}
normalized probability iMPS;
ordered stochastic MPO layers;
compression parameters and solver tolerances;
final evolution time $t_{\max}$ and measurement times
$\mathcal T_{\rm meas}$;
number of independent buffer samples $N_s$ per measurement;
number of blocks $N_b$ into which these samples are
partitioned for bootstrap estimation of the Markov length and
its sampling uncertainty;
maximum buffer length $L_{\max}$.

\If{$0\in\mathcal T_{\rm meas}$}
    \State Measure and save the initial-state CMI using the
    same procedure as below.
\EndIf

\For{$t=1,\ldots,t_{\max}$}
    \For{each layer in the model-specific ordering}
        \State Apply the MPO without intermediate truncation and compute dominant transfer-matrix eigenpairs $\langle\mathcal L|$ and $|\mathcal R\rangle$.
        \If{the expanded bond dimension exceeds
        $\chi_{\max}$}
            \State Solve Eqs.~\eqref{eq:sstein}, check
            convergence, and record the relative residuals.
            \State Construct the finite Hankel core $K$ in
            Eq.~\eqref{eq:shankel_core}, select the retained
            rank, and apply the truncation maps.
            \State Recompute the dominant transfer-matrix
            eigenpairs $\langle\mathcal L_\chi|$ and
            $|\mathcal R_\chi\rangle$, check their consistency, 
            and normalize the reduced tensors and environments.
        \EndIf
    \EndFor

    \If{$t\in\mathcal T_{\rm meas}$}
        \State Obtain the left and right fixed-point environments
        of the current iMPS transfer matrix.
        \State Sample $N_s$ spatial buffers over
        $1\leq L\leq L_{\max}$, with the probability-validity checks
        of Sec.~\ref{sec:cmi_statistics}.
        \State Partition the samples into $N_b$ blocks.
        \State Save the CMI means and sampling standard errors,
        together with each block's full CMI profile
        $(I_1,\ldots,I_{L_{\max}})$ for bootstrap estimation
        of the Markov length.
    \EndIf
\EndFor

\Statex \textbf{Output:}
final normalized probability iMPS;
CMI profiles and their sampling standard errors at all
measurement times;
block-level profiles $(I_1,\ldots,I_{L_{\max}})$ for bootstrap
estimation of the Markov length.
\end{algorithmic}
\end{algorithm}

\subsection{Evolution stability}
\paragraph{Rank-selection criterion.---}
After each MPO application, we compress the enlarged virtual
space by retaining the leading Hankel modes.
The retained rank is
\begin{equation}
    \chi
    =
    \min\!\left\{
        \chi_{\max},
        \#
        \left\{
            \alpha:
            \frac{\sigma_\alpha}{\sigma_1}
            \geq\epsilon_{\rm H}
        \right\}
    \right\}.
    \label{eq:snumericalrank}
\end{equation}
Here $\chi_{\max}$ caps the computational cost, while
$\epsilon_{\rm H}$ is a relative cutoff applied to the
singular values of the length-discounted Hankel matrix
$H_\beta$.
Discarding modes below this cutoff also stabilizes the
truncation maps in Eq.~\eqref{eq:struncationmaps}, whose
$\sigma_\alpha^{-1/2}$ factors would otherwise amplify
finite-precision errors associated with nearly vanishing
singular values.
The parameters used for DP and pcDP are listed in
Tables~\ref{tab:sdp_settings} and
\ref{tab:spcdp_settings}, respectively.

\paragraph{Numerical accuracy and stability.---}
We monitor numerical accuracy in constructing the truncation maps
and restore normalization after each evolution update.

We solve the Gramian equations in Eq.~\eqref{eq:sstein}
using matrix-free Krylov iteration, requiring solver convergence
and monitoring the relative residuals.
The solutions are symmetrized and factorized using positive
eigenmodes above a relative threshold
$\epsilon_{\rm supp}=10^{-15}$, measured against the largest
absolute eigenvalue.
The resulting factors determine the finite Hankel core and
the truncation maps in Eq.~\eqref{eq:struncationmaps}.

After each MPO application or compression, we compute the
dominant left and right eigenpairs of the transfer matrix.
We verify that the two dominant eigenvalues are real,
positive, and equal within numerical tolerance, and that the
corresponding eigenvectors have nonzero overlap.
We then divide the local tensors by the common dominant
eigenvalue, setting it to unity, and normalize the eigenvectors
such that $\langle\mathcal L|\mathcal R\rangle=1$.
The resulting boundary environments are used in the subsequent
truncation and probability-sampling steps.

\subsection{CMI measurements and statistical errors}
\label{sec:cmi_statistics}

At each measurement time, the sampler of Sec.~\ref{sec:sampling}
draws $N_s$ independent length-$L_{\max}$ spatial buffers from the
current normalized iMPS.
For each sampled trajectory, we require the conditional endpoint
probability $p(a,c\mid b)$ to have no entries below $-\varepsilon_{\rm neg}$, where $\varepsilon_{\rm neg}\simeq1.49\times10^{-8}$ is an absolute
numerical tolerance.
Negative entries within tolerance are set to zero, and the probability
is then renormalized to sum to unity.

For the length-$L$ prefix of sample $n$, we evaluate
\begin{equation}
Y_n(L)=I\!\left(A:C\mid B_L=\hat s_{1:L}^{(n)}\right).
\label{eq:ssample_contribution}
\end{equation}
Averaging over the independent samples gives the CMI estimate
and the pointwise sampling standard error used for the error
bars in the CMI-profile figures.
\begin{align}
\overline I_L&=\frac{1}{N_s}\sum_{n=1}^{N_s}Y_n(L),\notag\\
\operatorname{SE}_{\rm traj}(I_L)
&=\left[\frac{\sum_{n=1}^{N_s}(Y_n(L)-\overline I_L)^2}
{N_s(N_s-1)}\right]^{1/2}.
\label{eq:strajectory_error}
\end{align}
For the subsequent Markov-length fit and uncertainty analysis,
we need to account for correlations between CMI estimates
at different buffer lengths.
To preserve these correlations while reducing storage,
we partition the $N_s$ trajectories into $N_b$ disjoint
blocks of equal size $n_b=N_s/N_b$ and save the mean CMI
profile of each block:
\begin{equation}
    I_b(L)
    =
    \frac{1}{n_b}
    \sum_{n=(b-1)n_b+1}^{bn_b}Y_n(L),
    \qquad b=1,\ldots,N_b.
    \label{eq:scmi_blocks}
\end{equation}
Each profile
$I_b=\bigl(I_b(1),\ldots,I_b(L_{\max})\bigr)$
uses the same group of trajectories at every buffer length,
reducing the stored data from $N_s$ to $N_b$ profiles.
The overall mean and its sampling standard error are estimated
from the block profiles as
\begin{align}
    \overline I_L
    &=
    \frac{1}{N_b}\sum_{b=1}^{N_b}I_b(L),
    \notag\\
    \operatorname{SE}_{\rm block}(I_L)
    &=
    \left[
        \frac{1}{N_b(N_b-1)}
        \sum_{b=1}^{N_b}
        \bigl(I_b(L)-\overline I_L\bigr)^2
    \right]^{1/2}.
    \label{eq:sblock_error}
\end{align}
Equal block sizes ensure that $\overline I_L$ is identical
to the average over all trajectories.
These means and standard errors determine the weighted
Markov-length fit described below.
To estimate its sampling uncertainty, we bootstrap entire
block profiles, preserving the correlations across buffer lengths.

\subsection{Markov-length inference}
Algorithm~\ref{alg:mkl_inference} estimates the Markov length
and its sampling uncertainty from the saved block profiles.
At each measurement time, we fit the logarithm of the mean CMI
in a selected large-buffer window by weighted least squares
to obtain $\xi_{\rm M}$.
We estimate its uncertainty by resampling entire block profiles
with replacement and repeating the fit with the same window
and weights.

\begin{algorithm}[H]
\caption{Markov-length fit and block-bootstrap uncertainty.}
\label{alg:mkl_inference}
\small
\begin{algorithmic}[1]
\Statex \textbf{Input:}
block-level CMI profiles $I_b(L)$;
tail-window criteria;
number of bootstrap replicates $N_{\rm boot}$.

\For{each measurement time}
    \State Select a large-$L$ fitting window.
    \State Average all block profiles and compute their block
    standard errors.
    \State Perform the weighted fit in
    Eq.~\eqref{eq:sweighted} to obtain the central
    estimate $\xi_{\rm M}$.
    \For{$r=1,\ldots,N_{\rm boot}$}
        \State Draw $N_b$ complete block profiles uniformly
        with replacement and average them to construct a
        resampled CMI profile.
       \State Repeat the same weighted fit on the resampled
        profile, keeping the selected window and weights fixed, to
        obtain $\xi_{\rm M}^{(r)}$.
    \EndFor
    \State Report the central estimate $\xi_{\rm M}$ from
    the original mean CMI profile, together with the
    $16$th--$84$th percentile interval of the valid
    $\xi_{\rm M}^{(r)}$ estimates as its bootstrap sampling
    uncertainty.
\EndFor

\Statex \textbf{Output:}
central Markov lengths and bootstrap uncertainty intervals.
\end{algorithmic}
\end{algorithm}

The central estimate of $\xi_{\rm M}(t)$ is obtained by
fitting the mean CMI profile computed from all $N_b$ blocks.
On the selected window, we determine the
intercept and slope from
\begin{equation}
    (\widehat a,\widehat b)
    =
    \underset{a,b}{\operatorname{arg\,min}}
    \sum_{L\in\mathcal W(t)}
    w_L
    \left[
        \log\overline I_L-a-bL
    \right]^2,
    \qquad
    \xi_{\rm M}(t)
    =
    -\frac{1}{\widehat b},
    \label{eq:sweighted}
\end{equation}
with inverse-variance weights
$w_L=[\operatorname{SE}_{\rm block}(I_L)/\overline I_L]^{-2}$.

To estimate the sampling uncertainty of the fitted Markov
length, we bootstrap complete block-level CMI profiles,
thereby preserving correlations between different buffer
lengths.
For each of the $N_{\rm boot}$ replicates, we draw $N_b$
profiles with replacement from the original $N_b$ blocks,
average them to form a resampled CMI profile, and repeat the
weighted fit in Eq.~\eqref{eq:sweighted} using the same fitting
window and weights.
This produces a bootstrap estimate $\xi_{\rm M}^{(r)}$.
The shaded uncertainty bands show the 16th--84th percentiles
of the bootstrap fits.

\section{Directed-percolation benchmark}
\label{sec:dp_benchmark}
This section provides the model-specific implementation and numerical
settings underlying the directed-percolation benchmark in
Fig.~\ref{fig:comparison} of the main text.
We also test the sensitivity of the Hankel-based results to the
truncation parameters. The general numerical workflow is described
in Sec.~\ref{sec:numerical_details}.

\subsection{Model and stochastic MPO representation}
We use the Stavskaya directed-percolation dynamics
\cite{groverdp}. The model consists of a
one-dimensional chain of binary random variables $S_j$, with outcomes
$s_j=1$ denotes an active site and $s_j=0$ an inactive site. At even
(odd) time steps, only the odd (even) sublattice is updated.
Conditioned on the configuration at time $t$, the updated sites are
sampled independently according to
\begin{equation}
\Pr\!\left[
S_j(t+1)=0\mid
S_{j-1}(t)=s_{j-1},S_{j+1}(t)=s_{j+1}
\right]
=
\begin{cases}
1, & (s_{j-1},s_{j+1})=(0,0),\\
p, & \text{otherwise}.
\end{cases}
\label{eq:stavskaya_update}
\end{equation}
We use the critical-point parameter $p=0.295$~\cite{groverdp} and initialize the system in the fully active product state,
$s_j(0)\equiv1$ for every site.

\begin{figure}[h]
    \centering
    \includegraphics[width=\columnwidth]{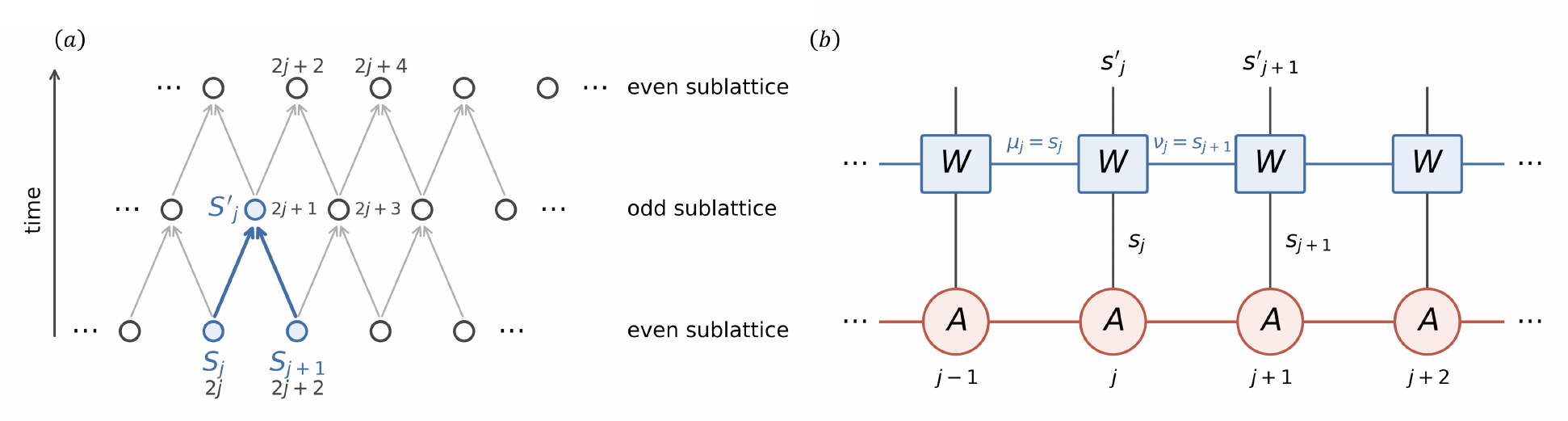}
   \caption{
    Staggered update and its tensor-network representation
    for the directed-percolation benchmark.
    (a) Successive updates alternate between the even and odd
    sublattices.The highlighted arrows identify the two parent
    random variables $S_j$ and $S_{j+1}$ of the updated random
    variable $S'_j$.
    (b) Application of the stochastic MPO to the probability iMPS.
    Blue squares denote the local update tensors $W$, and red
    circles denote the probability tensors $A$.
    The lower physical indices $s_j$ specify the input
    configuration, while the open upper indices $s'_j$ specify the
    updated configuration.
    The local delta constraint enforces $\mu_j=s_j$, while
    contraction of neighboring MPO tensors gives
    $\nu_j=\mu_{j+1}=s_{j+1}$.
    The tensor at site $j$ therefore implements the two-parent
    transition probability
    $q_p(s'_j\mid s_j,s_{j+1})$ in
    Eq.~\eqref{eq:dp_mpo_tensor}.
    }
    \label{fig:dp_update_schematic}
\end{figure}

In the numerical implementation, we relabel the sublattice updated
at each time step and represent its probability distribution as a
one-site translation-invariant iMPS.
The represented row consequently alternates between the odd and even
sublattices as time advances, as illustrated in
Fig.~\ref{fig:dp_update_schematic}(a).
In this representation, the next-row variable $s'_j$ has the
two parent variables $s_j$ and $s_{j+1}$.
Writing their conditional update probability as
$q_p(s_j'\mid s_j,s_{j+1})$, the local update tensor applying to the iMPS is
\begin{equation}
    W_{\mu\nu}^{s',s}
    =
    \delta_{\mu,s}\,
    q_p(s'\mid s,\nu),
    \qquad
    s,s',\mu,\nu\in\{0,1\}.
    \label{eq:dp_mpo_tensor}
\end{equation}
Here $\mu$ and $\nu$ denote the left and right MPO bond
indices. At site $j$, the Kronecker delta enforces
$\mu_j=s_j$. Contracting neighboring tensors identifies
$\nu_j=\mu_{j+1}$, and the delta constraint at site $j+1$
then gives $\nu_j=s_{j+1}$. Thus, the tensor at site $j$
contributes precisely
$q_p(s'_j\mid s_j,s_{j+1})$.
Contracting all local tensors therefore yields the
bond-dimension-$2$ stochastic MPO shown in
Fig.~\ref{fig:dp_update_schematic}(b), which implements the
full staggered-sublattice update rule.

\begin{table}[h]
\caption{
Numerical settings and achieved bond dimensions for the
directed-percolation benchmark. The first two rows define the primary
comparison in Fig.~\ref{fig:comparison} of the main text;
the remaining Hankel-based runs are
used for the parameter-sensitivity comparison in
Fig.~\ref{fig:dp-parameters}. Here $\chi_{\rm peak}$ is the largest bond
dimension retained during the evolution. The Hankel and Schmidt cutoffs
act on different spectra and are therefore not directly comparable.
}
\label{tab:sdp_settings}
\begin{ruledtabular}
\begin{tabular}{lcccc}
Run & $\beta$ & $\chi_{\max}$ & $\chi_{\rm peak}$ & Relative cutoff\\
\hline
Hankel benchmark & 0.990 & 180 & 180 & $10^{-14}$\\
canonical benchmark & -- & 180 & 104 & $4\times10^{-14}$\\
Hankel sensitivity & 0.985 & 160 & 160 & $10^{-14}$\\
Hankel sensitivity & 0.985 & 180 & 180 & $10^{-14}$\\
Hankel sensitivity & 0.995 & 180 & 180 & $4\times10^{-14}$
\end{tabular}
\end{ruledtabular}
\end{table}

\subsection{Benchmark protocol and numerical settings}
The numerical settings for the two methods compared in
Fig.~\ref{fig:comparison} are listed in the first two rows of
Table~\ref{tab:sdp_settings}.  Both runs choose $L_{max}=80$ and
$N_s = 10^4$ sampled trajectories. We partition the sampled trajectories 
into $N_b = 1000$ blocks and use bootstrap resampling 
to extract the Markov length and quantify its sampling uncertainty.

Although both calculations use the same bond-dimension cap
$\chi_{\max}=180$, only the Hankel-based evolution reaches this
cap. In the canonical calculation, the retained dimension is
selected automatically by the relative Schmidt-value cutoff
listed in Table~\ref{tab:sdp_settings}; it peaks at
$\chi_{\rm peak}=104$, with no additional restriction imposed
at that value.
In our Float64 implementation, lowering the cutoff to retain
additional Schmidt modes instead leads to earlier failures
of the probability-validity checks, with negative entries
exceeding the allowed tolerance.
This behavior is consistent with the amplification of numerical
errors when retaining modes with very small Schmidt values.
The smaller retained bond dimension of the canonical run
should therefore not be interpreted as greater compression
efficiency; rather, it is consistent with the mismatch between
Schmidt weight and the relevance of virtual directions to
probability-level observables.

Markov lengths are extracted using Algorithm~\ref{alg:mkl_inference}.
At each measurement time, we select a common fitting window
from the Hankel-based CMI profile.
Starting at the largest buffer length with a finite mean CMI
above $I_{\min}=10^{-14}$ and a finite, positive block standard
error, the window extends toward smaller $L$ over at most
$n_{\rm win}=30$ consecutive saved buffer lengths.
Both methods are fitted over this window, using their
respective block standard errors to determine the weights.

We estimate the sampling uncertainty using
$N_{\rm boot}=5000$ bootstrap replicates.
Each replicate resamples $N_b$ entire block profiles with
replacement and repeats the fit, keeping the selected buffer
lengths and the original fit weights fixed.
A replicate is valid only if all resampled CMI means at the
selected lengths are finite and positive and the fit yields
a finite, positive Markov length.
We report an estimate only if at least $95\%$ of the replicates
are valid.
The central value is obtained from the original mean profile,
and the uncertainty band spans the $16$th--$84$th percentiles
of the valid bootstrap estimates.

\subsection{Parameter sensitivity}
To assess sensitivity to the truncation parameters, we repeat
the DP evolution with different discount factors $\beta$,
bond-dimension caps $\chi_{\max}$, and relative Hankel
singular-value cutoffs.
The additional parameter sets are listed in
Table~\ref{tab:sdp_settings}.

\begin{figure}[t]
    \centering
    \includegraphics[width=\columnwidth]{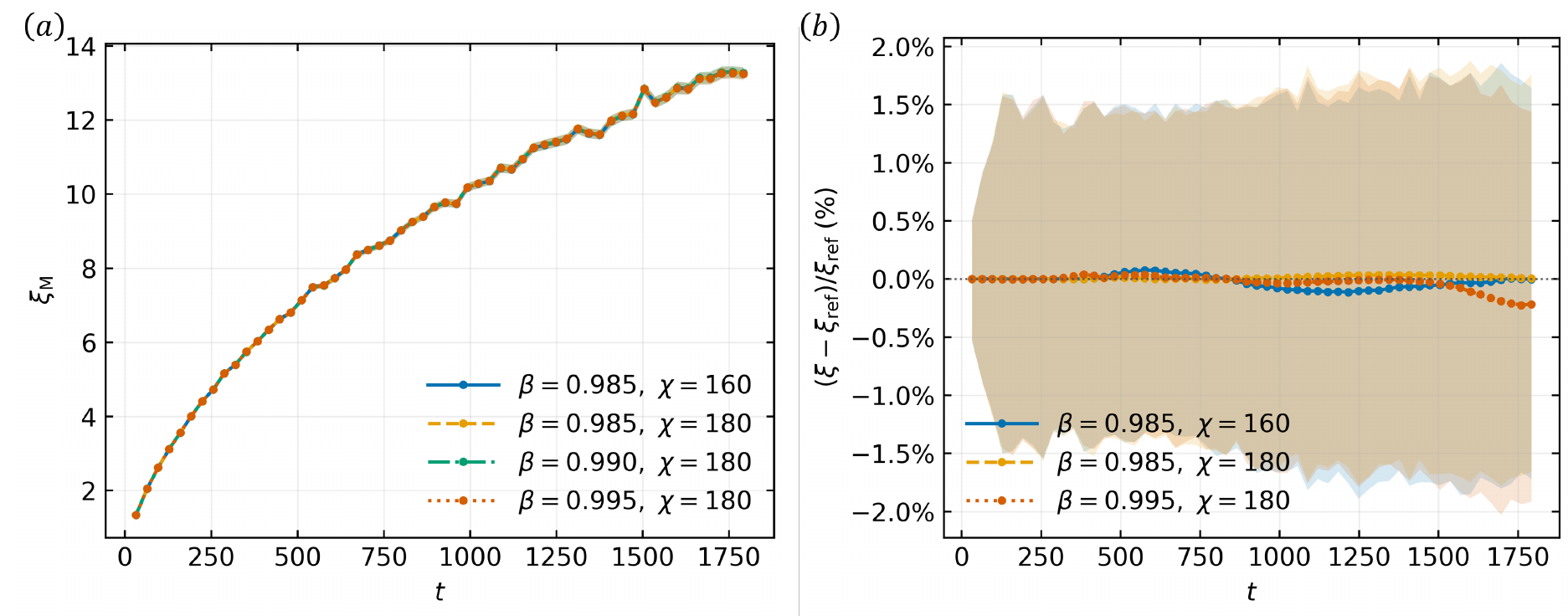}
   \caption{Parameter sensitivity of Hankel-based truncation in
    directed-percolation dynamics at $p=0.295$.
    (a) Markov lengths for the Hankel-based settings in
    Table~\ref{tab:sdp_settings}, with time measured in tilted-row updates.
    In the legends, $\chi$ denotes the bond-dimension cap $\chi_{\max}$.
    (b) Relative deviations
    $(\xi_{\rm M}-\xi_{\rm ref})/\xi_{\rm ref}$, shown as
    percentages, from the reference run with $\beta=0.990$
    and $\chi_{\max}=180$.
    Shaded bands indicate sampling uncertainty, showing the
    16th--84th percentiles of whole-block bootstrap estimates.
    The bands in (b) include the sampling uncertainty of both the
    compared and reference runs.}
    \label{fig:dp-parameters}
\end{figure}

Figure~\ref{fig:dp-parameters} compares the resulting Markov
lengths using the same fitting and bootstrap procedure.
At each time, the run with $\beta=0.990$ and $\chi_{\max}=180$
defines the fitting window shared by all runs.
Over the common resolved time range, the estimates agree
within their bootstrap uncertainties, supporting the numerical
reliability of the extracted Markov lengths.

\section{pcDP numerical implementation and validation}
\label{sec:pcdp_protocol}
This section details the model-specific stochastic MPO
construction and numerical settings used for the pcDP results
in Fig.~\ref{fig:pcdp_collapse} of the main text.
We also describe the direct Monte Carlo simulations used to
bracket the active--absorbing transition and validate the
domain-wall density obtained from the probability iMPS.
The general numerical workflow is described in
Sec.~\ref{sec:numerical_details}.

\subsection{Stochastic MPO construction and exact bond reduction}
The pcDP evolution is organized into the following four-layer
Floquet protocol:
\begin{equation}
    \mathcal F
    = \mathcal B_{\rm even}(Q)\mathcal B_{\rm odd}(Q)
      \mathcal D_{\rm even}(P,R)\mathcal D_{\rm odd}(P,R).
    \label{eq:pcdp_floquet}
\end{equation}
These layers implement the diffusion, pair-annihilation, and
branching processes defined in Eq.~\eqref{eq:pcdp_updates}.
We now construct the local update tensors that form each
stochastic MPO layer.

We first construct the diffusion--annihilation layers
$\mathcal D_{\rm odd}$ and $\mathcal D_{\rm even}$.
Let $s_{\rm L},s,s_{\rm R}\in\{\pm1\}$ denote the input spins
on the left neighboring site, the updated site, and the right
neighboring site, respectively.
The central spin flips with probability
\begin{equation}
    p_{\rm f}(s_{\rm L},s,s_{\rm R})
    =
    \begin{cases}
        0,
        & s_{\rm L}=s=s_{\rm R},\\
        P,
        & s_{\rm L}=s\ne s_{\rm R}
          \ \text{or}\ 
          s_{\rm L}\ne s=s_{\rm R},\\
        R,
        & s_{\rm L}=s_{\rm R}\ne s.
    \end{cases}
    \label{eq:spcdp_site_flip_probability}
\end{equation}

Because the flip probability
$p_{\rm f}(s_{\rm L},s,s_{\rm R})$ depends on the two
neighboring input spins, the MPO must transmit their values to
the updated site through its virtual bonds.
We implement this by alternating an update tensor $D$ on the
updated sublattice with a copy tensor $\Delta$ on the
complementary unupdated sublattice.
The binary virtual indices
$\mu,\nu\in\{\pm1\}$ carry the input spins from the left and
right neighboring sites, respectively.

Let $s$ and $s'$ denote the input and output spins at an
updated site.
The update tensor is
\begin{equation}
    \bigl[D^{s',s}\bigr]_{\mu\nu}
    =
    \begin{cases}
        1-p_{\rm f}(\mu,s,\nu),
        & s'=s,\\
        p_{\rm f}(\mu,s,\nu),
        & s'=-s,\\
        0,
        & \text{otherwise}.
    \end{cases}
    \label{eq:spcdp_site_mpo_tensor}
\end{equation}
An unupdated site is represented by the copy tensor
\begin{equation}
    \Delta^{s',s}_{\mu\nu}
    =
    \delta_{s',s}\delta_{\mu,s}\delta_{\nu,s},
\end{equation}
which preserves the physical spin and transmits its input
value to both adjacent virtual bonds.
Contracting alternating $D$ and $\Delta$ tensors therefore
sets $\mu=s_{\rm L}$ and $\nu=s_{\rm R}$ in $D$, reproducing
the local transition rule.
Placing $D$ on the odd or even sublattice yields
$\mathcal D_{\rm odd}$ or $\mathcal D_{\rm even}$,
respectively, with $\Delta$ acting on the complementary
sublattice.

The branching layers require a larger auxiliary space because
they implement a four-site rule.
Let $Y_j=(a_j,b_j)$ denote the spin pair being updated
in a given branching layer and
$Y'_j=(a'_j,b'_j)$ its output configuration.
The two spins flip together with probability $Q$ when
$b_{j-1}=a_j\ne b_j=a_{j+1}$, where $b_{j-1}$ and $a_{j+1}$
are the neighboring spins immediately outside the pair.
We denote the corresponding conditional transition probability by
\begin{equation}
    w_Q(a'_j,b'_j\mid b_{j-1},a_j,b_j,a_{j+1})
    =
    \begin{cases}
        Q,
        &
        \begin{gathered}
            b_{j-1}=a_j\ne b_j=a_{j+1},\qquad(a'_j,b'_j)=(-a_j,-b_j),
        \end{gathered}\\
        1-Q,
        &
        \begin{gathered}
            b_{j-1}=a_j\ne b_j=a_{j+1},\qquad(a'_j,b'_j)=(a_j,b_j),
        \end{gathered}\\
        1,
        &
        \begin{gathered}
            b_{j-1}=a_j\ne b_j=a_{j+1}
            \ \text{is not satisfied},\qquad (a'_j,b'_j)=(a_j,b_j),
        \end{gathered}\\
        0,
        & \text{all other output pairs}.
    \end{cases}
    \label{eq:spcdp_branch_weight}
\end{equation}

To encode the four-site dependence as a nearest-neighbor MPO,
we introduce two types of auxiliary bond indices,
$\alpha_j$ and $\beta_j$:
\begin{align}
    \alpha_j
    &=
    (b_{j-1},a_j,a'_j),
    & \dim\alpha&=2^3=8,
    \notag\\
    \beta_j
    &=
    (\alpha_j,b_j,b'_j)
    =
    (b_{j-1},a_j,a'_j,b_j,b'_j),
    & \dim\beta&=2^5=32.
    \label{eq:spcdp_branch_memory}
\end{align}
Here $\alpha_j$ stores the left boundary spin and the
input--output pair on the first site of block $j$, while
$\beta_j$ additionally stores the pair on its second site.
This encoding represents the branching layer as an alternating
sequence of tensors $T_a$ and $T_b$, placed on the left and
right sites of each updated block, respectively.
The tensor $T_b$ at site $b_j$ appends the local input--output
pair $(b_j,b'_j)$ to $\alpha_j$, forming $\beta_j$.
The tensor $T_a$ at site $a_{j+1}$ then combines this stored
information with the right neighboring input spin $a_{j+1}$
to evaluate the transition weight $w_Q$ for updated pair $Y_j$.
It also passes the information needed for the next block
through $\alpha_{j+1}=(b_j,a_{j+1},a'_{j+1})$.
Explicitly,
\begin{align}
    \bigl[T_b^{b'_j,b_j}\bigr]_{\alpha_j,\beta_j}
    &=
    \begin{cases}
        1,
        & \beta_j=(\alpha_j,b_j,b'_j),\\
        0,
        & \text{otherwise},
    \end{cases}
    \notag\\
    \bigl[T_a^{a'_{j+1},a_{j+1}}\bigr]
    _{\beta_j,\alpha_{j+1}}
    &=
    \begin{cases}
        w_Q\!\left(
            a'_j,b'_j
            \mid
            b_{j-1},a_j,b_j,a_{j+1}
        \right),
        &
        \alpha_{j+1}
        =
        (b_j,a_{j+1},a'_{j+1}),\\
        0,
        & \text{otherwise}.
    \end{cases}
    \label{eq:spcdp_raw_branch_tensors}
\end{align}

\begin{figure}[t]
    \centering
    \includegraphics[width=0.92\columnwidth]{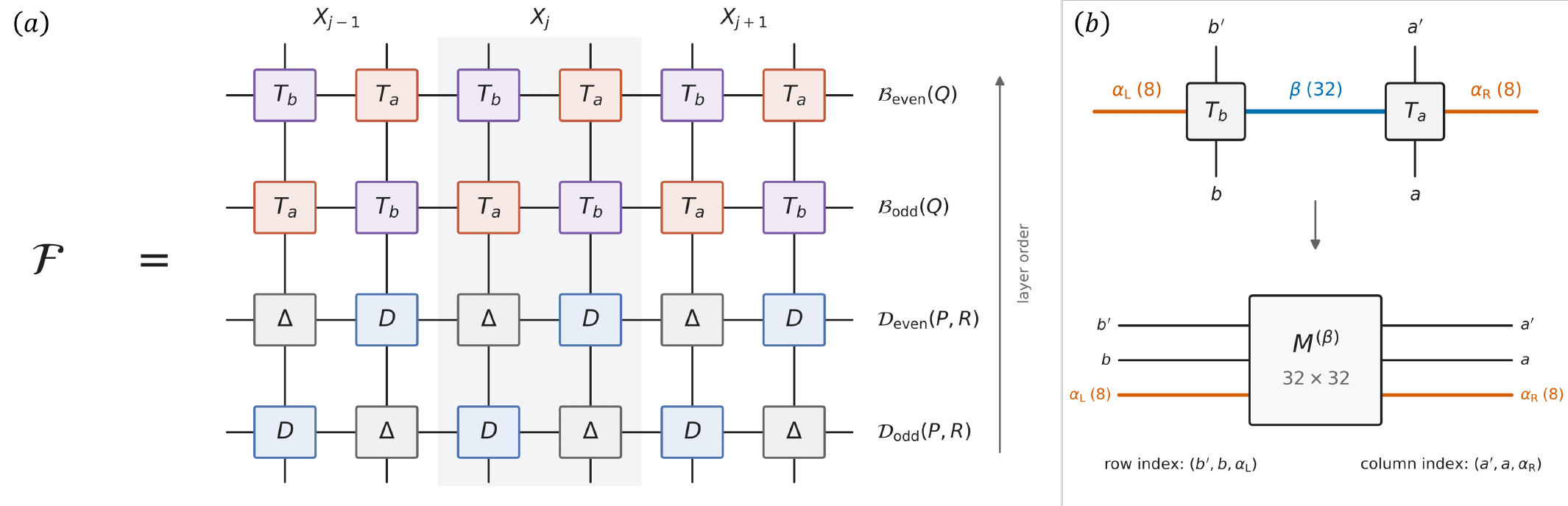}
    \caption{
    Stochastic MPO representation and exact bond reduction for
    pcDP. (a) Four-layer tensor network for the Floquet period in
    Eq.~\eqref{eq:pcdp_floquet}. The layers act from bottom to top;
    $D$ and $\Delta$ alternate in the diffusion--annihilation
    layers, while $T_a$ and $T_b$ alternate in the branching
    layers. Horizontal legs are MPO bond indices, and vertical
    legs connect physical input and output indices between layers.
    (b)  Exact reduction of the $\beta$ bond
    in Eq.~\eqref{eq:spcdp_branch_memory}, illustrated for the
    $\mathcal B_{\rm odd}$ layer.
    Contracting $T_b$ and $T_a$
    across $\beta$ and grouping the remaining indices gives the
    $32\times32$ matrix $M^{(\beta)}$, whose row and column indices
    are $(b',b,\alpha_{\rm L})$ and
    $(a',a,\alpha_{\rm R})$, respectively. Merging proportional
    columns according to Eq.~\eqref{eq:spcdp_proportional_columns}
    reduces $\dim\beta$ from $32$ to $4$ while preserving the
    two-tensor contraction exactly.
    }
    \label{fig:spcdp_mpo_bond_reduction}
\end{figure}

Contracting the alternating tensors over all auxiliary indices
yields the product of conditional transition weights
$\prod_j w_Q$, reproducing the full branching update.
Following the convention in the main text,
$\mathcal B_{\rm odd}$ updates the pairs $(2j-1,2j)$,
with $T_a$ and $T_b$ placed on odd and even sites,
respectively.
Exchanging their positions gives $\mathcal B_{\rm even}$,
which updates the pairs $(2j,2j+1)$.
Together with the diffusion--annihilation layers, these MPOs
form the four-layer Floquet network shown in
Fig.~\ref{fig:spcdp_mpo_bond_reduction}(a). For the numerical iMPS representation, we use the fixed two-site cells $X_j=(S_{2j-1},S_{2j})$ and retain this partition throughout all four layers. After grouping the two local MPO tensors associated with each
cell, every layer is represented as an MPO acting on the
blocked chain.
Its bond dimension $\kappa$ is the dimension of the remaining
auxiliary index connecting neighboring cells.

\paragraph{Exact bond reduction.---}
Although construction in Eq.\eqref{eq:spcdp_raw_branch_tensors} is exact, 
its auxiliary spaces contain substantial redundancy: 
distinct virtual states can produce proportional contributions 
to the same neighboring tensor contraction. 
We remove this redundancy once, before time evolution.

For each alternating bond, we contract the two neighboring
tensors and reshape the result into a nonnegative matrix $M$,
as illustrated for the $\beta$ bond in
Fig.~\ref{fig:spcdp_mpo_bond_reduction}(b).
For each nonzero column $\boldsymbol m_k$, we define
\begin{equation}
    c_k
    =
    \boldsymbol 1^T\boldsymbol m_k,
    \qquad
    \widehat{\boldsymbol m}_k
    =
    \frac{\boldsymbol m_k}{c_k}.
    \label{eq:spcdp_proportional_columns}
\end{equation}
Columns with the same normalized profile
$\widehat{\boldsymbol m}_k$ are proportional and are therefore
merged into a single auxiliary state, with the scale $c_k$
absorbed into the neighboring tensor.
This nonnegative reduction preserves each two-tensor
contraction exactly, while compressing the auxiliary spaces from
$\dim\beta=32$ and $\dim\alpha=8$ to
$\dim\widetilde\beta=4$ and
$\dim\widetilde\alpha=5$, respectively.
After two-site blocking, the resulting branching MPOs
$\mathcal B_{\rm odd}$ and $\mathcal B_{\rm even}$ therefore
have bond dimensions $\kappa=4$ and $5$, respectively.

\subsection{Parameter Choice and probability-iMPS validation}
\paragraph{Transition scan.---}
To bracket the active--absorbing transition, we perform direct
Monte Carlo (MC) simulations of the four-layer Floquet
dynamics in Eq.~\eqref{eq:pcdp_floquet}, using the same
unbiased product initial state as in the probability-iMPS
calculation,
$\Pr(S_i=+1)=\Pr(S_i=-1)=1/2$.
We fix $P=0.50$ and $R=0.30$ and scan
$Q=0.120$, $0.125$, $0.130$, $0.135$, and $0.140$.
For each $Q$, we evolve $N_{\rm traj}=10^4$ independent
trajectories of a periodic chain with
$L_{\rm MC}=2^{15}$ sites up to
$t_{\max}^{\rm MC}=1440$ complete Floquet periods.

We use the domain-wall density as the order parameter for the
active--absorbing transition.
Although the Floquet protocol distinguishes intra-cell
bonds from inter-cell bonds, their densities agree within
the numerical accuracy and time windows considered.
We therefore set $\rho(t)\equiv\rho_{\rm AB}(t)$.
The MC estimator is
\begin{equation}
    \rho_{\rm intra}^{\rm MC}(t;Q)
    =
    \frac{2}{L_{\rm MC}}
    \sum_{j=1}^{L_{\rm MC}/2}
    \left\langle
        \frac{
            1-S_{2j-1}(t)S_{2j}(t)
        }{2}
    \right\rangle_{\rm MC}.
    \label{eq:spcdp_density_ab_mc}
\end{equation}
For the translation-invariant probability iMPS, the same
observable is evaluated as
\begin{equation}
    \rho_{\rm intra}^{\rm iMPS}(t;Q)
    =
    \sum_{a,b=\pm1}
    \frac{1-ab}{2}\,
    \langle\mathcal L_t|
    A_t^{a,b}
    |\mathcal R_t\rangle,
    \label{eq:spcdp_density_ab_imps}
\end{equation}
where $A_t^{a,b}$ is the iMPS tensor for spin values
$a$ and $b$ at sites $2j-1$ and $2j$, respectively. The density $\rho(t)$ used to rescale the CMI in Fig.~\ref{fig:pcdp_collapse}(a) of the main text is
calculated from Eq.~\eqref{eq:spcdp_density_ab_imps}.

\begin{figure*}[t]
    \centering
    \includegraphics[width=\textwidth]{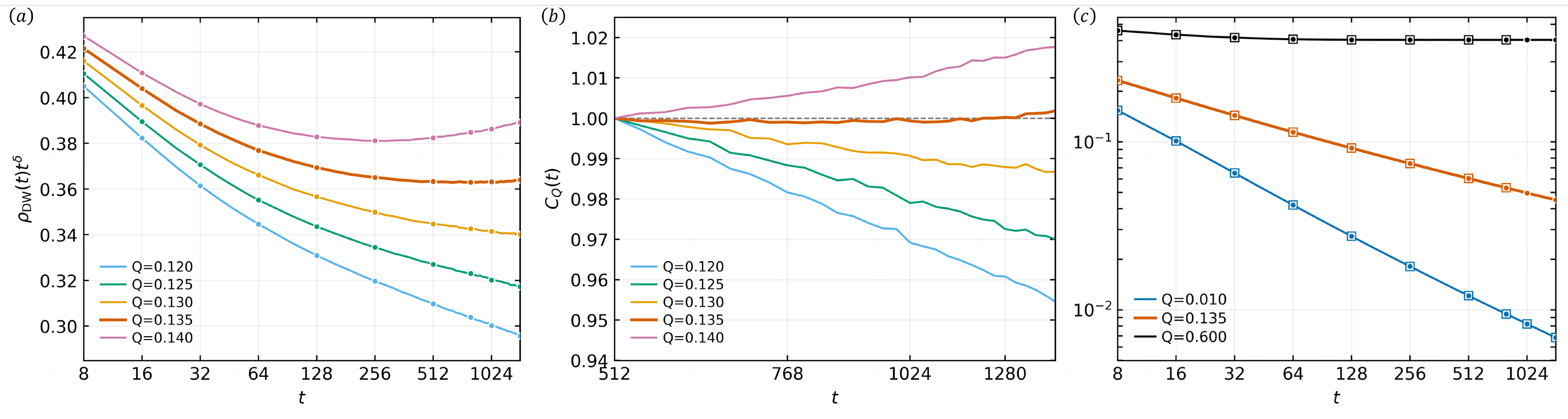}
    \caption{
    Finite-time MC scan and probability-iMPS density benchmark for pcDP.
    (a) Compensated domain-wall density
    $\rho(t;Q)t^\delta$ with $\delta=0.28719$.
    Markers indicate selected times; error bars show one
    standard error and are smaller than the symbols where not
    visible.
    (b) Compensated density normalized at $t_0=512$ according
    to Eq.~\eqref{eq:spcdp_compensated_density}; the dashed
    line marks $C_Q(t)=1$.
    (c) Domain-wall density at the inactive ($Q=0.010$),
    critical ($Q=0.135$), and active ($Q=0.600$) regimes. 
    Solid curves and filled circles show direct MC results,
    while open squares show probability-iMPS results. MC standard
    errors are smaller than the symbols.
    }
    \label{fig:spcdp-validation}
\end{figure*}

Figure~\ref{fig:spcdp-validation}(a) shows the compensated
density $\rho(t;Q)t^\delta$, using the pcDP decay exponent
$\delta=0.28719$~\cite{park2013highprecision}.
At criticality, this quantity approaches a constant at late
times.
To expose the residual drift, we normalize each curve at
$t_0=512$,
\begin{equation}
    C_Q(t)
    =\frac{\rho(t;Q)t^\delta}{\rho(t_0;Q)t_0^\delta},
    \qquad t_0=512.
    \label{eq:spcdp_compensated_density}
\end{equation}
As shown in Fig.~\ref{fig:spcdp-validation}(b),
$C_Q(t)$ drifts downward for $Q\leq0.130$ and upward for
$Q=0.140$, while the $Q=0.135$ curve remains nearly constant,
changing by only about $0.2\%$ up to $t=1440$.
We therefore take
$0.130\lesssim Q_c\lesssim0.140$ as a conservative finite-time
bracket and choose $Q=0.135$ as the critical point.
We also choose $Q=0.010$ and $0.600$ well inside the inactive and
active regimes, respectively. Numerical settings are summarized in Table~\ref{tab:spcdp_settings}.

\begin{table}[h]
    \caption{Numerical settings for the pcDP results shown in the figures
    of the main text. Each CMI data set uses $N_s=10^4$ sampled trajectories,
    $N_b = 1000$ sampling blocks, and $L_{max} = 60$ for the active phase while $L_{max} = 150$ for critical point and the inactive phase.}
    \label{tab:spcdp_settings}
    \begin{ruledtabular}
    \begin{tabular}{lccccc}
    Regime & $Q$ & $\chi_{\max}$ & $t_{\max}$ & $\beta$ & $\epsilon_{\rm H}$\\
    \hline
    Active & 0.600 & 420 & 800 & 0.99 & $10^{-12}$\\
    Critical & 0.135 & 800 & 800 & 0.995 & $10^{-13}$\\
    Inactive & 0.010 & 400 & 1440 & 0.995 & $10^{-13}$
    \end{tabular}
    \end{ruledtabular}
\end{table}

\paragraph{Density benchmark.---}
To validate the probability-iMPS evolution, we compare its
domain-wall density with direct MC results at the
representative inactive, critical, and active parameters
$Q=0.010$, $0.135$, and $0.600$, respectively.
As shown in Fig.~\ref{fig:spcdp-validation}(c), the iMPS
densities closely track the MC results throughout the
accessible time windows.
Across all matched observation times, the relative deviation
satisfies
$|\rho_{\rm AB}^{\rm iMPS}-\rho_{\rm AB}^{\rm MC}|
/\rho_{\rm AB}^{\rm MC} < 1.6\times10^{-3}$.

\section{Transverse-contraction calculation of temporal CMI in SSEP}
\label{sec:ssep_temporal_numerics}
This section details the numerical procedure used to calculate
the temporal CMI shown in
Fig.~\ref{fig:pcdp_collapse}(c) of the main text. The method
combines transverse contraction of the SSEP spacetime tensor
network with direct Monte Carlo sampling of the conditioning
records. We first construct the spacetime tensor network and
recast it in a form suitable for contraction along the spatial
direction. We then explain how the Monte Carlo samples are
combined with the tensor-network contraction to obtain the
conditional endpoint distributions entering the CMI. Finally,
we verify convergence with the transverse spatial radius and
specify the numerical parameters used for the reported results.

\subsection{SSEP spacetime tensor network}
\begin{figure*}[t]
    \centering
    \includegraphics[width=\textwidth]{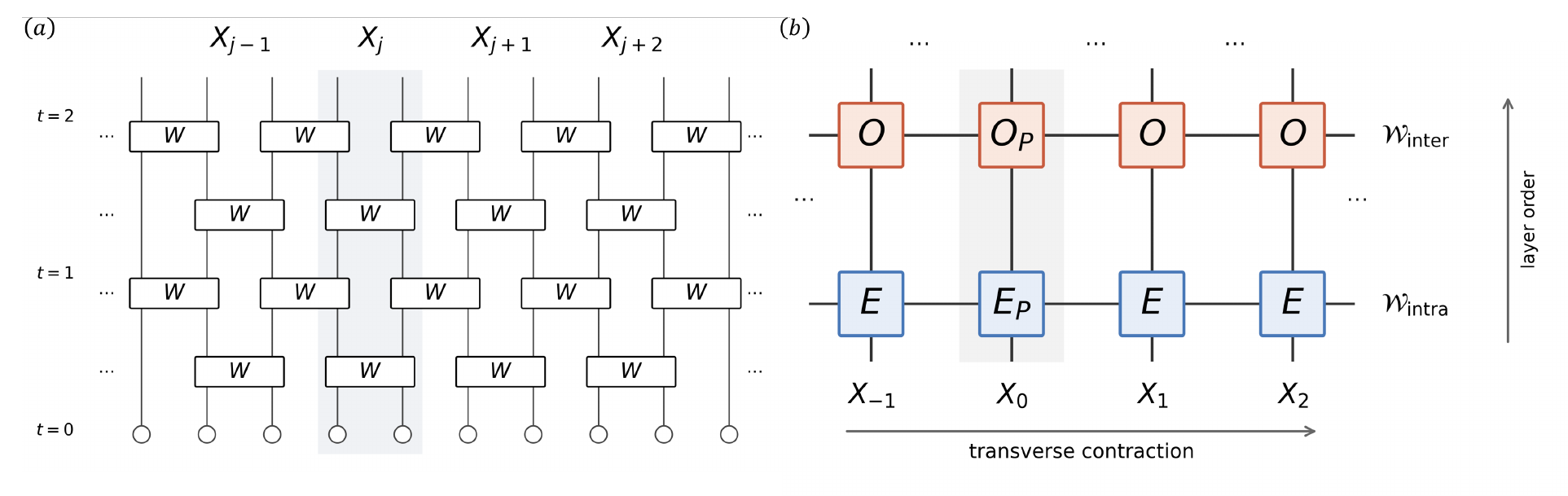}
    \caption{
    SSEP spacetime tensor network and transverse-contraction
    representation. (a) Site-resolved brickwork network generated
    by alternating the two-site stochastic gate $W$ on intra-cell
    and inter-cell bonds. The circles at $t=0$ denote the Bernoulli
    product-state, and the shaded region illustrates the
    two-site blocking into cells $X_j$. (b) Blocked MPO
    representation of one Floquet period,
    $\mathcal F_{\rm SSEP}=\mathcal W_{\rm inter}
    \mathcal W_{\rm intra}$. The lower and upper rows contain the
    intra-cell tensors $E$ and inter-cell tensors $O$, respectively
    [Eqs.~\eqref{eq:ssep_intra_tensor} and
    \eqref{eq:ssep_inter_tensor}]. The shaded column marks the probe block $X_0$, which contains the observed site $S_0$. The tensors $E_P$ and $O_P$ denote the corresponding probe-column tensors with 
    local occupation projectors inserted along the temporal worldline.
    }
    \label{fig:ssep_transverse_schematic}
\end{figure*}

As described in the main text, the spacetime tensor network is
initialized with the translation-invariant Bernoulli product
measure at half filling, in which each site is independently
occupied with probability $P=1/2$.
To exploit the two-site translation invariance of the staggered
Floquet dynamics, we group neighboring occupation variables into
two-site cells, $X_j=(S_{2j},S_{2j+1})$, with the ordered basis
$(00,01,10,11)$. Let $X=(s_1,s_2)$ and $X'=(s'_1,s'_2)$ denote the input
and output configurations of an updated bond, respectively, the corresponding two-site stochastic tensor is
\begin{equation}
W^{S'_1S'_2}_{S_1S_2}
\equiv W_{X',X},\qquad
W=\begin{pmatrix}
1&0&0&0\\
0&\frac12&\frac12&0\\
0&\frac12&\frac12&0\\
0&0&0&1
\end{pmatrix}.
\label{eq:ssep_local_gate}
\end{equation}
One complete Floquet period consists of two staggered layers.
The first layer updates the bond inside every cell,
$(2j,2j+1)$, and the second updates the bonds connecting
neighboring cells, $(2j+1,2j+2)$. Denoting these layers by
$\mathcal W_{\rm intra}$ and $\mathcal W_{\rm inter}$,
respectively, the one-period evolution operator is
\begin{equation}
\mathcal F_{\rm SSEP} = \mathcal W_{\rm inter}\mathcal W_{\rm intra}.
\label{eq:sssep_floquet}
\end{equation}
Repeating
Eq.~\eqref{eq:sssep_floquet} generates the spacetime tensor
network shown in Fig.~\ref{fig:ssep_transverse_schematic}(a).

Our goal is to evaluate the temporal CMI defined in the maintext.
For a fixed conditioning record, the occupations observed at
the probe site $S_0$ are imposed by inserting local occupation
projectors along its temporal
worldline~\cite{iadecola2026local}.
These projectors single out the probe block as a measured
defect column, while all unmeasured spatial blocks remain
identical by translation invariance.
We therefore contract the network transversely, along the
spatial direction: the tensors within each spatial block are
grouped into a temporal MPO, and repeated applications of the
common bulk temporal MPO propagate the left and right temporal
boundary states toward the defect column.
Reading upward along the time direction, the bulk temporal MPO
has the schematic alternating structure
$\operatorname{tMPO}[E,O,E,O,\ldots]$, where $E$ and $O$ are
the intra-cell and inter-cell tensors, respectively
[Fig.~\ref{fig:ssep_transverse_schematic}(b)].
The intra-cell gate is obtained directly from $W$,
\begin{equation}
E^{(s'_{2j},s'_{2j+1}),(s_{2j},s_{2j+1})} = W^{s'_{2j},s'_{2j+1}}_{s_{2j},s_{2j+1}}.
\label{eq:ssep_intra_tensor}
\end{equation}
The inter-cell gate, however, requires an exact decomposition into a sum of four local product Pauli terms:
\begin{equation}
    \begin{aligned}
        W&=\sum_{\mu\in\{0,x,y,z\}}M_\mu\otimes M_\mu, \\
        O^{(s'_{2j},s'_{2j+1}),(s_{2j},s_{2j+1})}_{\mu_{j-1},\mu_j}
        &=
        \left(M_{\mu_{j-1}}\right)^{s'_{2j}}_{s_{2j}}
        \left(M_{\mu_j}\right)^{s'_{2j+1}}_{s_{2j+1}}.
    \end{aligned}
    \label{eq:ssep_inter_tensor}
\end{equation}
Here
$(M_0,M_x,M_y,M_z)=\frac12(\sqrt3 I,X,Y,Z)$, while
$\mu_j\in\{0,x,y,z\}$ is the four-dimensional MPO bond between
blocks $X_j$ and $X_{j+1}$. With this construction, the
spacetime tensor network containing the observation projectors
is recast in the transverse form shown in
Fig.~\ref{fig:ssep_transverse_schematic}(b).

Numerically, we initialize the left and right temporal boundary
MPS at the natural open spatial boundaries and propagate them
toward the defect column by successively applying the
unmeasured temporal MPO. After each application, the boundary MPS
is compressed along the temporal direction using canonical SVD
truncation with cutoff $\epsilon_{\rm SVD}$ and maximum
bond dimension $\chi_{\max}$. We increase the transverse
radius $R$ until the resulting CMI becomes insensitive to
further increases in $R$, indicating convergence to the
thermodynamic-limit value within numerical accuracy. 
The Monte Carlo sampling of the conditioning records and the
corresponding tensor-network evaluation of the CMI are described
below.

\subsection{Monte Carlo sampling of temporal CMI}
Directly summing over all possible conditioning records requires
enumerating a number of binary strings that grows exponentially
with the temporal separation. We instead sample the intervening
records from independent classical SSEP trajectories uing the same Floquet dynamics on a sufficiently large periodic chain with
$L_{\rm MC}=65536$. The initial configuration of each trajectory
is drawn uniformly from the fixed-particle-number sector with
$N=L_{\rm MC}/2$. 
Stationarity ensures that the joint distribution of a
contiguous temporal record depends only on its length and not on
the absolute starting time. Moreover, the fixed-number ensemble
used in the Monte Carlo simulation and the Bernoulli ensemble
used in the tensor-network contraction have the same local
spacetime marginals in the thermodynamic limit. The Monte Carlo records can therefore be used directly as the conditioning records in the tensor-network evaluation of the temporal CMI.

Concretely, along each trajectory we record the occupation of
the probe site $S_0$ after every complete Floquet period up to
the endpoint time $t$. The $q$th trajectory produces the
conditioning record
\begin{equation}
\boldsymbol b^{(q)}(t)
=
\bigl(
s^{(q)}_0(1),s^{(q)}_0(2),\ldots,s^{(q)}_0(t-1)
\bigr),
\qquad q=1,\ldots,N_s .
\label{eq:ssep_sampled_record}
\end{equation}
At each intermediate time $\tau$, the recorded occupation
$s^{(q)}_0(\tau)\in\{0,1\}$ selects the local projector
\begin{equation}
\Pi_{s^{(q)}_0(\tau)}
=
\sum_{x=0}^{1}
\bigl|s^{(q)}_0(\tau),x\bigr\rangle
\bigl\langle s^{(q)}_0(\tau),x\bigr|,
\qquad \tau=1,\ldots,t-1.
\label{eq:ssep_sampled_projector}
\end{equation}
Here $x$ denotes the occupation of the unobserved right site of
the probe block $X_0$. The projector is inserted on the
corresponding temporal link of the probe worldline.

We next construct the measured defect MPOs. Evaluating the CMI
requires the conditional endpoint distribution
$p(S_0(0),S_0(t)\mid\boldsymbol b^{(q)})$. Thereofre, for each sampled record, we consider all four endpoint choices and insert the corresponding projectors $\Pi_{S_0(0)}$ and $\Pi_{S_0(t)}$ at the lower
and upper ends of the probe worldline. The resulting four defect MPOs are denoted by $\mathbb T^{(q)}_{a,c}$, where $a \equiv S_0(0)$ and $c \equiv S_0(t)$.

For each sampled record $\boldsymbol b^{(q)}$, contracting the
four defect MPOs with the left and right temporal boundary
environments gives the unnormalized endpoint weights
\begin{equation}
w^{(q)}_{a,c} =
\langle\mathcal L_R|
\mathbb T^{(q)}_{a,c}
|\mathcal R_R\rangle,
\qquad a,c\in\{0,1\}.
\label{eq:ssep_endpoint_weights}
\end{equation}
Normalizing these four weights gives the conditional endpoint distribution
\begin{equation}
p_R\!\left(a,c\mid\boldsymbol b^{(q)}\right)
=
\frac{w^{(q)}_{a,c}}
{\displaystyle\sum_{a',c'}w^{(q)}_{a',c'}}.
\label{eq:ssep_conditional_endpoint}
\end{equation}
From this distribution, we evaluate the record-resolved mutual
information $I_R(A:C\mid B=\boldsymbol b^{(q)})$ using $\sum_{a,c} p_R(a,c\mid \boldsymbol b^{(q)} )
    \ln\frac{p_R(a,c\mid \boldsymbol b^{(q)} )}
    {p_R(a\mid \boldsymbol b^{(q)} )p_R(c\mid \boldsymbol b^{(q)})}$.The temporal CMI is then obtained by
substituting these values into the Monte Carlo estimator
Eq.~\eqref{eq:sestimator}.

\subsection{Numerical convergence and production parameters}
\begin{figure}[h]
    \centering
    \includegraphics[width=0.8\columnwidth]{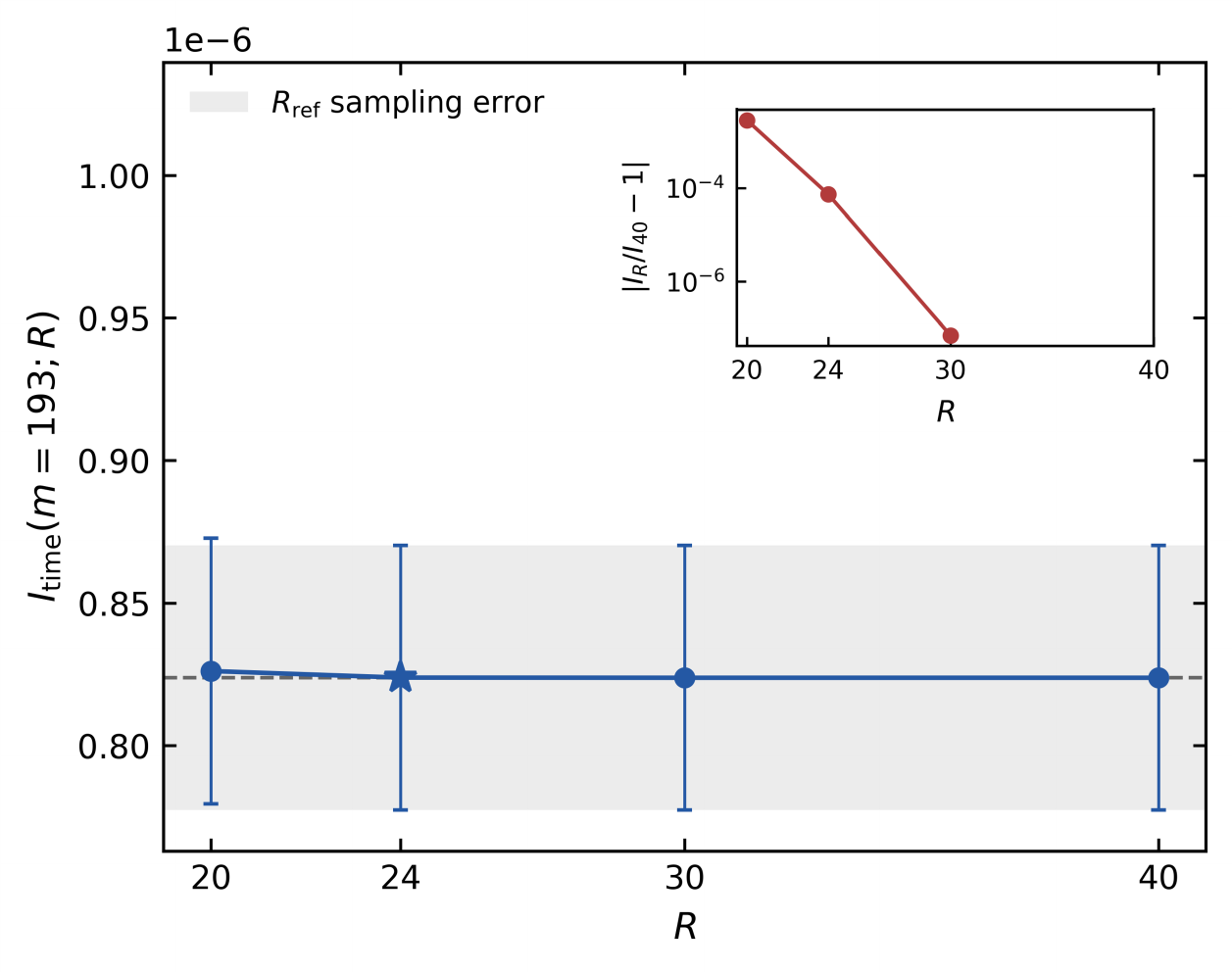}
    \caption{
    Convergence of the temporal CMI with the transverse spatial
    radius $R$ at the largest endpoint separation, $t=193$.
    The dashed line gives the $R=40$ reference value,
    and the shaded region denotes its sampling standard error.
    The star marks the production radius $R=24$. The inset shows
    the relative deviation $|I_R/I_{40}-1|$ on a logarithmic
    scale.
    }
    \label{fig:sssep_radius_convergence}
\end{figure}

We now assess the finite-radius error at the largest endpoint
separation considered in the main text, $t=193$, where the
temporal boundary states are most costly to represent. 
To this end, we evaluate $R=20,24,30,$ and $40$ using the same
set of $32$ conditioning records, so that differences between
the resulting CMI estimates directly probe the finite-radius
dependence. Figure~\ref{fig:sssep_radius_convergence} shows the
sample-averaged temporal CMI, with $R=40$ taken as the reference.
The results are converged within sampling uncertainty already at
$R=24$. We therefore use
\begin{equation}
R=24,\qquad
\chi_{\max}=256,\qquad
\epsilon_{\rm SVD}=10^{-12}
\label{eq:sssep_production_parameters}
\end{equation}
for the calculations reported in the main text.

\end{document}